%% file: main.tex
\documentclass[10pt,twoside,twocolumn,a4paper,journal]{IEEEtran}
\usepackage{cite}
\usepackage[utf8]{inputenc}
\usepackage{multirow}
\usepackage{amsmath}
\usepackage{amsfonts}
\usepackage[ruled,linesnumbered]{algorithm2e}
\usepackage{commands,color}
\usepackage{tabularx}
\usepackage{url}
\usepackage{gensymb}
\usepackage{graphicx}
\usepackage{subcaption}

\newcommand{\redalert}[1]{{\color{red}#1}}

\title{Auxiliary Function-Based Algorithm for Blind Extraction of a Moving Speaker}

\author{
Jakub Jansk\'y, Zbyn\v{e}k Koldovsk\'{y}, Ji\v{r}\'i M\'alek, Tom\'a\v{s} Kounovsk\'y, and Jaroslav \v{C}mejla
\vspace{0.1in} \\
Acoustic Signal Analysis and Processing Group, Faculty of 
Mechatronics, 
Informatics, and Interdisciplinary\\ Studies,
Technical University of Liberec, Studentsk\'a 2, 461 17
Liberec, Czech Republic. \\E-mail:
zbynek.koldovsky@tul.cz, fax:+420-485-353112, tel:+420-485-353534
}

\begin{document}

\maketitle

\footnotetext{This work was supported by The Czech Science Foundation through 
Projects No.~17-00902S and No.~20-17720S, by the United States Department of the Navy, Office of Naval Research Global, through Project No.~N62909-19-1-2105 and by the Student Grant Competition of the Technical University of Liberec under the project No. SGS-2019-3060.
}
	
\begin{abstract}
Recently, Constant Separating Vector (CSV) mixing model has been proposed for 
the Blind Source Extraction (BSE) of moving sources. In this paper, we 
experimentally verify the applicability of CSV in the blind extraction of a 
moving speaker and propose a new BSE method derived by modifying the auxiliary 
function-based algorithm for Independent Vector Analysis. Also, a piloted 
variant is proposed for the method with partially controllable global 
convergence. The methods are verified under reverberant and noisy conditions 
using { simulated as well as real-world acoustic conditions}. They are also 
verified within the CHiME-4 speech separation and recognition challenge. The 
experiments corroborate the applicability of CSV as well as the improved 
convergence of the proposed algorithms.
\end{abstract}

\section{Introduction}
A mixture of audio signals that propagate in an acoustic environment from point sources to microphones can be described by the convolutive model. Let there be $d$ sources observed by $m$ microphones. The signal on the $i$th microphone is described by
\begin{equation}\label{eq:model}
    x_i(n)=\sum_{j=1}^d\sum_{\tau=0}^{L-1}
    h_{ij}(\tau)s_j(n-\tau),\quad i=1,\dots,m,
\end{equation}
where $n$ is the sample index, $s_1(n),\dots,s_d(n)$ are the original signals coming from the sources, and $h_{ij}$ denotes the impulse response between the $j$th source and $i$th microphone of length $L$. 

In the Short-Time Fourier Transform (STFT) domain, convolution can be approximated by multiplication. Let $x_i(k,\ell)$ and $s_j(k,\ell)$ denote, respectively, the STFT coefficient of $x_i(n)$ and $s_j(n)$ at frequency $k$ and frame $\ell$. Then,  \eqref{eq:model} can be replaced by a set of $K$ complex-valued instantaneous mixtures
\begin{equation}\label{eq:STFT_BSS_model}
    {\bf x}_k={\bf A}_k {\bf s}_k, \qquad k = 1,\dots,K,
\end{equation}
where ${\bf x}_k$ and ${\bf s}_k$ are symbolic vectors representing, respectively,  $[x_1(k,\ell),\dots,x_r(k,\ell)]^T$ and $[s_1(k,\ell),\dots,s_d(k,\ell)]^T$, for any frame $\ell$; ${\bf A}_k$ stands for the $m\times d$ mixing matrix whose $ij$th element is related to the $k$th Fourier coefficient of the impulse response $h_{ij}$; $K$ is the frequency resolution of the STFT; for detailed explanations, see, e.g., Chapters 1 through 3 in \cite{ASSSEbook2018}.

This paper addresses the multi-channel speech extraction problem where the goal is to extract a selected speech signal from the microphone recordings. We particularly address the problem when the corresponding speaker is moving.  Unknown situation is considered where no information about the mixing parameters and the positions of microphones and sources is available. Solutions through Blind Source Separation (BSS), particularly through Blind Source Extraction (BSE), are sought. These methods are mostly based on mathematical models and assumptions and provide alternatives to learning-based approaches that require large data and computational resources for training \cite{wang2018}.

The mixing parameters in \eqref{eq:model} and \eqref{eq:STFT_BSS_model}, that is, the impulse responses and the mixing matrices, respectively,
are constant over $n$ and $\ell$. The models can thus describe only situations where sources and acoustic conditions are static. Recordings involving moving sources can be modeled by making these mixing parameters time or frame-dependent \cite{taseska2018}. We will consider the processing of signals in the STFT domain.

BSS embraces a large number of methods, vast majority of which assume static 
mixing models \cite{makino2007,comon2010handbook}. Independent Component 
Analysis (ICA), the most popular one, { can separate {\em statistically 
independent} signals up to their order and scales 
\cite{comon1994,hyvarinen2001}}. Its deployment in audio applications in the 
STFT domain is called Frequency-Domain ICA (FDICA). In FDICA, each mixture in 
\eqref{eq:STFT_BSS_model}, that is, each frequency bin, is processed separately 
\cite{smaragdis1998,ASSSEbook2018}. This gives rise to the permutation problem: 
The separated frequency components have random order and must be aligned in 
order to retrieve the full-band separated signals \cite{sawada2004sap}. 
Independent Vector Analysis (IVA) is an extension of ICA that treats all 
mixtures in \eqref{eq:STFT_BSS_model} simultaneously using a joint statistical 
model \cite{kim2006,kim2007,ono2011stable}. The frequency components of the 
original signals form the so-called vector components, and IVA aims at 
maximizing their mutual higher-order dependencies while the whole vector 
components should be independent \cite{kim2007}.
A recent extension of IVA is Independent Low Rank Matrix Analysis (ILRMA)  where the vector components are assumed to obey a low-rank model, which is actually an advanced vector-source model. For example, ILRMA combines the IVA and Nonnegative Matrix Factorization (NMF) in \cite{kitamura2016determined,kitamura2018}.

The above static mixture-based methods can be used to separate/extract moving sources by being applied on short intervals where the mixture is approximately static. Indeed, on-line recursive modifications are typically implemented to process data sample-by-sample or frame-by-frame using an exponential forgetting update of inner parameters; see, e.g., \cite{taniguchi2014,khan2015}. The tuning of such on-line implementations poses a difficult and application-dependent problem. 

In this paper, we consider an alternative BSE approach that contributes to the 
solution of the moving speaker problem. It is based on a new mixing model 
called the Constant Separating Vector (CSV)  model, which allows for changes of 
the mixing parameters within a given batch of data. The model helps to avoid 
the problem of extracting different sources at different times, referred to as 
the {\em discontinuity problem}, which is crucial in on-line BSE. Also, the 
corresponding achievable extraction accuracy given by the Cram\'er-Rao bound 
for this model is, in theory, higher than that of static methods when applied 
to short intervals \cite{kautsky2019CRLB}. These advantages are payed for by 
the following limitation: CSV inherently assumes that the desired signal can be 
extracted from the given mixture by a constant beamformer steered towards the 
area of the target source occurrence during the movement { 
\cite{koldovsky2020fastdiva}}. We address this property in this paper and show 
that its practical impact on the applicability of CSV { in the moving speaker 
extraction problem is not critical}. 

CSV has been recently introduced in the theoretical study in 
\cite{kautsky2019CRLB} as a special case of piecewise determined mixing models. 
In \cite{koldovsky2019icassp}, its deployment in a speech extraction experiment 
has been briefly mentioned. This paper provides the missing information in 
\cite{koldovsky2019icassp}, that is, a complete definition of CSV and the 
details of its application in the speaker extraction problem. Also, a detailed 
derivation of the gradient-based algorithm, BOGIVE$_{\bf w}$, used in 
\cite{koldovsky2019icassp} is given. { The main contribution resides in the 
proposal of} an auxiliary-function based algorithm (Block~AuxIVE), which is 
much faster and less prone to local extremes compared to BOGIVE$_{\bf w}$. 
{Also}, a piloted version of Block~AuxIVE is proposed that features a partially 
controlled global convergence through a pilot signal \cite{nesta2017supervised}.

The article is organized as follows. In Section~II, the problem of the blind extraction of a moving speaker based on the CSV mixing model is formulated. In Section~III, the proposed algorithms BOGIVE$_{\bf w}$, Block AuxIVE and piloted Block~AuxIVE are derived. Section~IV is devoted to experimental evaluations based on simulated as well as real-world data. The paper is concluded in Section~V.

\paragraph*{Notation} Plain letters denote scalars, bold letters denote vectors, and bold capital letters denote matrices. Upper indices such as $\cdot^T$, 
$\cdot^H$, or $\cdot^*$ denote, respectively, transposition, conjugate transpose, or complex conjugate. The Matlab convention for matrix/vector concatenation and 
indexing will be used, e.g., $[1;\,{\bf g}]=[1,\, {\bf g}^T]^T$ 
and $({\bf a})_i$ is the $i$th element of ${\bf a}$. ${\rm E}[\cdot]$ stands for the expectation operator, and $\hat{\rm E}[\cdot]$ is the average taken over all available samples of the symbolic argument. The letters $k$ and $t$ are used as integer indices of frequency bin and block, respectively; $\{\cdot\}_k$ is a short notation of the argument with all values of index $k$, e.g., $\{{\bf w}_k\}_k$ means ${\bf w}_1,\dots,{\bf w}_K$, and $\{{\bf w}_{k,t}\}_{k,t}$ means ${\bf w}_{1,1},\dots,{\bf w}_{K,T}$.

\section{Problem Formulation}


\subsection{Blind Source Extraction}In BSE, the goal is to extract a particular 
{\em source of interest} (SOI) from the observed signals. Owing to the 
ambiguity of {order and scales of the original signals}, it is not possible to 
guarantee the extraction of the SOI without additional information (e.g., by a 
proper initialization of an algorithm). {Any independent signal in the observed 
mixture} can play the role of the SOI. We will get back to the problem of 
controlling the convergence to the desired SOI in Section~\ref{sec:piloted} and 
in the experimental part of the paper.

{
The BSS mixing model in the STFT domain given by \eqref{eq:STFT_BSS_model} can be written in the form
\begin{equation}\label{eq:BSE_mixmodel}
    {\bf x}_k = {\bf a}_k s_k + {\bf y}_k,\qquad k=1,\dots,K,
\end{equation}
where $s_k$ represents the SOI, which corresponds to the first original signal in ${\bf s}_k$, ${\bf a}_k$ is the first column of ${\bf A}_k$, called the {\em mixing vector}; ${\bf y}_k$ represents  the remaining signals. The assumption that \eqref{eq:BSE_mixmodel} is a rewrite of \eqref{eq:STFT_BSS_model} where the mixing matrices ${\bf A}_k$ are square and non-singular, hence invertible, guarantees the existence of a {\em separating vector} ${\bf w}_k$ such that ${\bf w}_k^H{\bf x}_k=s_k$. It brings the limitation that ${\bf y}_k$ involves (only) $d-1$ linearly mixed signals\footnote{The existence of ${\bf w}_k$ is important for the development of efficient algorithms; the assumption can be violated to some extent when applied in practice, as will be shown in the experimental section.}. 

The main assumption, similar to ICA and IVA, is that $s_k$ and ${\bf y}_k$ are independent. Compared to ICA/IVA, ${\bf y}_k$ need not necessarily be the mixture of independent signals as its separation into the original signals is not the goal of BSE. 

In Independent Component/Vector Extraction (ICE/IVE), recently introduced in \cite{koldovsky2019TSP}\footnote{ICE and IVE were shown to be the subproblems of ICA and IVA, closely related to the previous nonGaussianity-based approaches \cite{hyvarinen1999,delfosse1995,lee2007fast}.}, a reduced parameterization of the mixing and de-mixing matrix is assumed, which involves only the parameters related to the the SOI: the mixing and the separating vector.} It was proven that this structure is sufficient for the BSE task.
Specifically, the mixing matrix in \eqref{eq:STFT_BSS_model} and its inverse matrix can be parameterized, respectively, as
\begin{equation}\label{eq:mixingmatrix}
   \Aal = 
   \begin{pmatrix}
\Al &     {\bf Q}_k
\end{pmatrix}  = 
 \begin{pmatrix}
  \gamma_k & {\bf h}_k^H\\
   {\bf g}_k &   \frac{1}{\gamma_k}({\bf g}_k{\bf h}_k^H-\I_{d-1}) \\
    \end{pmatrix},
\end{equation}
 and
\begin{equation}\label{eq:demixingmatrix}
    \Wl = 
     \begin{pmatrix}
     {\bf w}_k^H\\
     \Bl
     \end{pmatrix}  = 
     \begin{pmatrix}
     {\beta_k}^* & {\bf h}_k^H\\
     {\bf g}_k & -\gamma_k  \I_{d-1} \\
     \end{pmatrix},
\end{equation}
where $\I_d$ denotes the $d \times d$ identity matrix, ${\bf w}_k$ denotes the separating vector which is partitioned as ${\bf w}_k=[\beta_k;{\bf h}_k]$;  the mixing vector ${\bf a}_k$ is partitioned as ${\bf a}_k=[\gamma_k;{\bf g}_k]$. The vectors ${\bf a}_k$ and ${\bf w}_k$ are linked through the so-called {\em distortionless constraint} ${\bf w}_k^H{\bf a}_k = 1$, which, equivalently, means
\begin{equation}\label{eq:distortionlesscst}
    \beta_k^*\gamma_k + {\bf h}_k^H{\bf g}_k = 1, \qquad k=1,\dots,K.
\end{equation}
${\bf B}_k=[{\bf g}_k, -\gamma_k  \I_{d-1}]$ is called {\em blocking matrix} as it satisfies that ${\bf B}_k{\bf a}_k={\bf 0}$. 

Let ${\bf z}_k={\bf B}_k{\bf x}_k={\bf B}_k{\bf y}_k$ be called the background 
signals. The mixing model ensures that ${\bf y}_k={\bf Q}_k{\bf z}_k$, so 
\eqref{eq:STFT_BSS_model} {as well as \eqref{eq:BSE_mixmodel}}, with the 
assumed structures \eqref{eq:mixingmatrix} and \eqref{eq:demixingmatrix}, can 
be written as
\begin{equation}\label{eq:IVEmodel}
    {\bf x}_k = \begin{pmatrix}
  \gamma_k & {\bf h}_k^H\\
   {\bf g}_k &   \frac{1}{\gamma_k}({\bf g}_k{\bf h}_k^H-\I_{d-1}) \\
    \end{pmatrix}
    \begin{pmatrix}
     s_k\\
     {\bf z}_k
     \end{pmatrix},
\end{equation}
$k=1,\dots,K$.

\subsection{CSV Mixing Model}
Now, we adapt the extension of \eqref{eq:IVEmodel} to time-varying mixtures. Let the available samples of the observed signals (meaning the STFT coefficients from $N$ frames) be divided into $T$ intervals; for the sake of simplicity, we assume that the intervals have the same integer length $N_b=N/T$. The intervals will be called blocks and will be indexed by $t\in \mathcal{T}=\{1,\dots,T\}$. 

A conventional extension of \eqref{eq:IVEmodel} to time-varying mixtures is when all parameters of the (de-)mixing matrices, i.e., the mixing and separating vectors, are block-dependent. In the Constant Separating Vector (CSV) mixing model, it is assumed that only the mixing vectors are block-dependent while the separating vectors are constant over the blocks. Specifically, the mixing  and de-mixing matrices are parameterized, respectively, as
\begin{equation}\label{eq:CSVmixingmatrix}
 {\bf A}_{k,t} = 
\begin{pmatrix}
{\bf a}_{k,t} & {\bf Q}_{k,t}
\end{pmatrix}  = 
 \begin{pmatrix}
  \gamma_{k,t} & {\bf h}_k^H\\
   {\bf g}_{k,t} &   \frac{1}{\gamma_{k,t}}({\bf g}_{k,t}{\bf h}_{k}^H-\I_{d-1}) \\
    \end{pmatrix},   
\end{equation}
and
\begin{equation}
 {\bf W}_{k,t} = 
 \begin{pmatrix}
 {\bf w}_{k}^H\\
 {\bf B}_{k,t}
 \end{pmatrix}  = 
 \begin{pmatrix}
{ {\beta_k^*}} & {\bf h}_k^H\\
  {\bf g}_{k,t} & -\gamma_{k,t}  \I_{d-1} \\
 \end{pmatrix}.
\end{equation}
Each sample of the observed signals on the $t$th block is modeled according to
\begin{equation}\label{eq:CSVmodel}
    {\bf x}_{k,t}={\bf A}_{k,t}\begin{pmatrix}
     s_{k,t}\\
     {\bf z}_{k,t}
     \end{pmatrix},
\end{equation}
where $s_{k,t}$ and ${\bf z}_{k,t}$ represent, respectively, the $k$th frequency of the SOI and of the background signals at any frame within the $t$th block. Obviously, the CSV coincides with the static model \eqref{eq:IVEmodel} when $T=1$.

\begin{figure}
    \centering
    \includegraphics[width=0.95\linewidth]{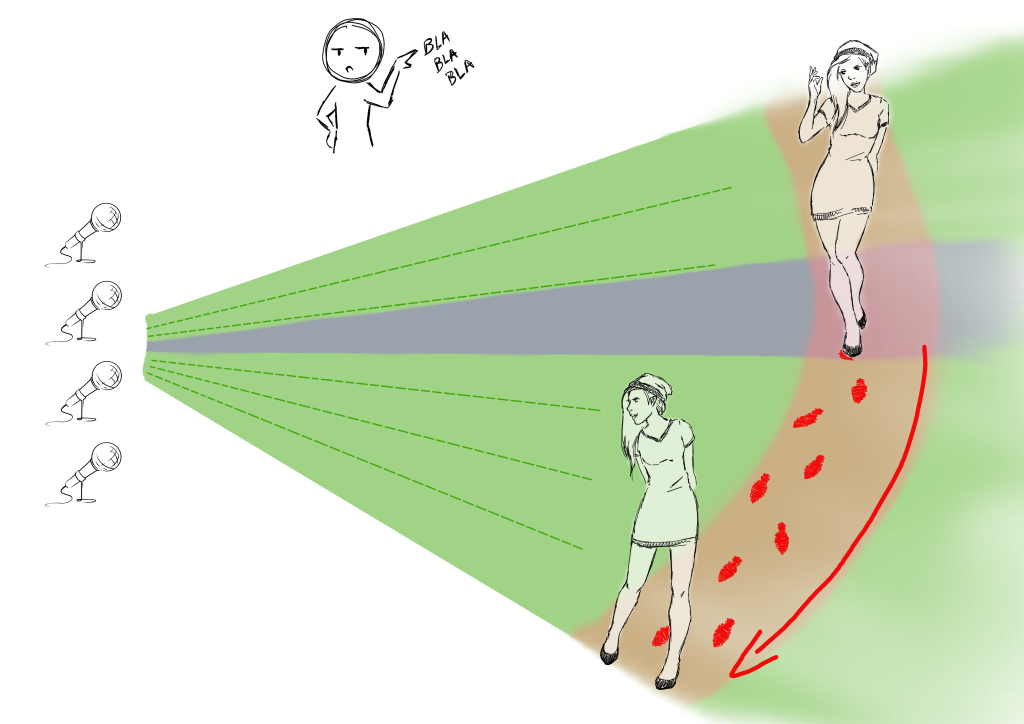}
    \caption{An illustration of how the blind extraction of a moving speaker can be solved based on CSV. The narrow area (in grey) stands for a typical focus of a separating filter obtained by the static mixing models. It is able to extract the speaker only from a particular position. The green area denotes the focus of a separating filter obtained through CSV: it covers the entire area of the speaker's movement during the recording. Such separating vector exists because there is a sufficient number of microphones and because the interfering speaker is located outside the area of the target speaker.}
    \label{fig:CSVillustration}
\end{figure}

The idea behind the CSV model is illustrated in Fig.~\ref{fig:CSVillustration}. While CSV admits that the SOI can change its position from block to block (the mixing vectors ${\bf a}_{k,t}$ depend on $t$), the block-independent separating vector is sought such that extracts the speaker's voice from all positions visited during its movement. There are two main reasons for this: First, the reduced number of mixing model parameters  increases the achievable interference-to-signal ratio to order $\mathcal{O}(N^{-1})$, compared to that of the conventional time-varying model, which is $\mathcal{O}(N_b^{-1})$; this is confirmed by the theoretical study on Cram\'er-Rao bounds in \cite{kautsky2019CRLB}. Second, the CSV enables BSE methods to avoid the discontinuity problem, which appears with the conventional model because each block is processed independently and different sources can be extracted in different blocks.

The CSV also brings some limitations. Formally, the mixture must obey the condition that for each $k$ a separating vector exists such that $s_{k,t}={\bf w}_{k}^H{\bf x}_{k,t}$ holds for every $t$; a condition that seems to be quite restrictive. Nevertheless, preliminary experiments in \cite{koldovsky2019icassp} have shown that CSV is useful in practical situations, especially when the number of microphones is high enough to provide sufficient degrees of freedom. The experimental part of this work presented in Section~IV validates these findings. 


\subsection{Source model}\label{sec:statisticalmodelstd}
Now, we introduce the statistical model of the signals. Let the vector component corresponding to the SOI be denoted by ${\bf s}_t=[s_{1,t},\dots,s_{K,t}]^T$. The elements of ${\bf s}_t$ are allowed to be dependent but uncorrelated, which is the idea adopted from IVE \cite{koldovsky2019TSP} and originating from IVA \cite{kim2007}. 

Let $p_s({\bf s}_t)$ denote the joint pdf of ${\bf s}_t$ and $p_{{\bf z}_{k,t}}({\bf z}_{k,t})$  denote the pdf\footnote{We might consider a joint pdf of ${\bf z}_{1,t},\dots,{\bf z}_{K,t}$ that could possibly involve higher-order dependencies between the background components. However, since $p_{{\bf z}_{k,t}}(\cdot)$ is assumed Gaussian in this paper, and since signals from different mixtures (frequencies) are assumed to be uncorrelated, as in the standard IVA, we can directly consider ${\bf z}_{1,t},\dots,{\bf z}_{K,t}$ to be mutually independent.} of ${\bf z}_{k,t}$. For simplicity, $p_s(\cdot)$ is denoted without the index $t$ but it is generally dependent on $t$. 
Since ${\bf s}_t$ and ${\bf z}_{1,t},\dots,{\bf z}_{K,t}$ are independent, their joint pdf within the $t$th block is equal to
\begin{equation}\label{eq:jointoriginalpdf}
    p_s({\bf s}_t)\cdot\prod_{k=1}^K p_{{\bf z}_{k,t}}({\bf z}_{k,t}).
\end{equation}

From \eqref{eq:CSVmodel} and \eqref{eq:jointoriginalpdf} it follows that the joint pdf of the observed signals in the $t$th block reads
\begin{multline}\label{eq:independence2}
   p_{\bf x}(\{{\bf x}_{k,t}\}_k) = p_s(\{{\bf w}_k^H{\bf x}_{k,t}\}_k)\times\\ \prod_{k=1}^K p_{{\bf z}_{k,t}}({\bf B}_{k,t}{\bf x}_{k,t}) |\det {\bf W}_{k,t}|^2.
\end{multline}
Hence, the log-likelihood function for one sample (frame) of the observed signals in the $t$th block is given by
\begin{multline}\label{eq:logogive_w}
    \mathcal{L}(\{{\bf w}_k\}_k,\{{\bf a}_{k,t}\}_k|\{{\bf x}_{k,t}\}_k) =\log p_s(\{{\hat s}_{k,t}\}_k)  \\ +\sum_{k=1}^{K} \log p_{{\bf z}_{k,t}}(\hat{\bf z}_{k,t}) +\log |\det {\bf W}_{k,t}|^2,
\end{multline}
where ${\hat s}_{k,t}={\bf w}_k^H{\bf x}_{k,t}$ and $\hat{\bf z}_{k,t}={\bf B}_{k,t}{\bf x}_{k,t}$ stand for the current estimate of the SOI and the background signals, respectively.

In BSS and BSE, the true pdfs of the original sources are not known, so suitable model densities have to be chosen. The rule of thumb says that the mismatch between the true and model densities has an influence on the separation/extraction accuracy \cite{fortunati2017}. Therefore, the aim is to select model densities that reflect the true properties of the source signals as much as possible. In BSE, it is typical to assume that the background signals are Gaussian as these are not subject to extraction \cite{koldovsky2019TSP}. The concrete choice of the model pdf for SOI, which must be non-Gaussian, will be discussed in Section~\ref{sec:nonlinearity}.

Let $f(\cdot)$ be the model pdf corresponding to a normalized non-Gaussian random variable. 
In ICA and IVA (and ICE and IVE), where data are processed in a single block due to the static mixing model, $p_s({\bf s}_t)$ can be replaced by $f({\bf s}_t)$. This is because the scaling ambiguity (the original scale of the SOI cannot be retrieved in BSE) enables us to extract the SOI with variance equal to one \cite{kim2007}. That is why $f(\cdot)$ can be normalized to unit variance. By contrast, in CSV, the changing variance of the SOI across the blocks has to be taken into account, otherwise the pdf modeling is not sufficiently accurate. Therefore, according to the probability density transformation theorem, the pdf on the $t$th block should be replaced by
\begin{equation}\label{eq:ps}
    p_s({\bf s}_t) \approx f\left(\left\{\frac{{ s}_{k,t}}{{\sigma}_{k,t}}\right\}_k\right)\left(\prod_{k=1}^K{\sigma}_{k,t}\right)^{-2},
\end{equation}
where $\sigma^2_{k,t}$ is the variance of $s_{k,t}$.

Now, we can derive a contrast function for the estimation of the model parameters by using the model pdfs in \eqref{eq:logogive_w}.
The pdf of the background will be assumed to be circular Gaussian with zero mean and (unknown) covariance matrix ${\bf C}_{{\bf z}_{k,t}}={\rm E}[{\bf z}_{k,t}{\bf z}_{k,t}^H]$, i.e., $p_{{\bf z}_{k,t}}\sim\mathcal{CN}(0,{\bf C}_{{\bf z}_{k,t}})$. The unknown ${\bf C}_{{\bf z}_{k,t}}$ is replaced by its sample-based estimate $\widehat{\bf C}_{{\bf z}_{k,t}}=\hat{\rm E}[\hat{\bf z}_{k,t}\hat{\bf z}_{k,t}^H]$. Similarly, $\sigma^2_{k,t}$ is replaced by the sample-based variance of $\hat s_{k,t}$, which is equal to $\hat\sigma_{k,t}=\sqrt{{\bf w}_k^H\widehat{\bf C}_{k,t}{\bf w}_k}$ where $\widehat{\bf C}_{k,t}=\hat{\rm E}[{\bf x}_{k,t}{\bf x}_{k,t}^H]$ is the sample-based covariance matrix of ${\bf x}_{k,t}$.

All signal samples are assumed to be independently distributed, so the instantaneous value of the log-quasilikelihood function can be replaced by averages over the blocks and available samples. Using $|\det {\bf W}_{k,t}|^2=|\gamma_{k,t}|^{2(d-2)}$ (follows from Eq. (15) in \cite{koldovsky2019TSP}), putting \eqref{eq:ps} into \eqref{eq:logogive_w}, and neglecting the constant terms, the contrast function obtains the form
\begin{multline}\label{eq:contastIVE}
    \mathcal{C}(\{{\bf w}_k\}_k,\{{\bf a}_{k,t}\}_{k,t}) =\\ \frac{1}{T}\sum_{t=1}^T\Bigl\{\hat{\rm E}\left[\log f\left(\left\{\frac{{\bf w}_k^H{\bf x}_{k,t}}{\hat\sigma_{k,t}}\right\}_k\right)\right] -\sum_{k=1}^{K}\log(\hat\sigma_{k,t})^2 \\ -\sum_{k=1}^{K} \hat{\rm E}[{\bf x}_{k,t}^H{\bf B}_{k,t}^H\widehat{\bf C}_{{\bf z}_{k,t}}^{-1}{\bf B}_{k,t}{\bf x}_{k,t}] +(d-2)\sum_{k=1}^{K}\log |\gamma_{k,t}|^2\Bigr\}.
\end{multline}
Note that $\widehat{\bf C}_{{\bf z}_{k,t}}$ is treated as a constant parameter in \eqref{eq:contastIVE}, while its value is updated after each update of ${\bf B}_{k,t}$, resp. ${\bf a}_{k,t}$.
In the following section, we propose numerical algorithms to find the maximum of \eqref{eq:contastIVE} subject to the separating and mixing vectors, which leads to their consistent estimation.


\section{Algorithms}

\subsection{Orthogonally Constrained Gradient Algorithm}\label{sec:OGIVEw}
The gradient algorithm proposed in this section will be referred to as 
BOGIVE$_{\bf w}$ (Block OrthoGonally constrained IVE), similarly to OGIVE$_{\bf w}$ proposed in \cite{koldovsky2019TSP}, because the former can be seen as an extension of the latter for $T>1$. BOGIVE$_{\bf w}$ iterates in small steps in the direction of a constrained gradient of \eqref{eq:contastIVE}. The constrained gradient is the gradient taken with respect to ${\bf w}_k$ when the mixing vectors ${\bf a}_{k,t}$, $t=1,\dots,T$, are dependent through the orthogonal constraint (OGC). The OGC must be imposed, because updating ${\bf a}_{k,t}$ and ${\bf w}_k$ as independent parameters (linked only through the distortionless constraint ${\bf a}_{k,t}{\bf w}_k = 1$) in the directions of unconstrained gradients has been shown to be highly unstable \cite{eusipco2017}.

The OGC ensures that the current estimate of the SOI ${\hat s}_{k,t}$ has zero sample correlation with the background signals $\hat{\bf z}_{k,t}$, that is, $\hat{\rm E}[{\hat s}_{k,t}\hat{\bf z}_{k,t}^H]={\bf w}_k^H\wCxkt{\bf B}_{k,t}={\bf 0}$, for every $k$ and $t$. From Appendix~A in  \cite{koldovsky2019TSP} it follows that the OGC is imposed when ${\bf a}_{k,t}$ depends on ${\bf w}_k$ through
\begin{equation}\label{eq:couplingaCSV}
	{\bf a}_{k,t}=\frac{\wCxkt{\bf w}_k}{{\bf w}_k^H\wCxkt{\bf w}_k}, \quad t\in\mathcal{T}.
\end{equation}
To specify, the constrained gradient of \eqref{eq:contastIVE} with respect to ${\bf w}_k$ is given by
\begin{equation}
    \Delta_{k} = \frac{\partial}{\partial {\bf w}_k^H} \mathcal{C}\left(\{{\bf w}_k\}_k,\left\{\frac{\wCxkt{\bf w}_k}{{\bf w}_k^H\wCxkt{\bf w}_k}\right\}_{k,t}\right).
\end{equation}

In order to find the expression for $\Delta_{k}$, we will use the following identities, which come from straightforward computations using the Wirtinger calculus \cite{kreutzdelgado2009}:
\begin{align}\label{eq:deriv_identity1a}
    \frac{\partial}{\partial {\bf w}_k}\frac{1}{\hat\sigma_{k,t}} & = - \frac{{\bf a}_{k,t}}{2\hat\sigma_{k,t}},\\\label{eq:deriv_identity1b}
    \frac{\partial}{\partial {\bf w}_k}\log \hat\sigma_{k,t}^2 & = {\bf a}_{k,t}.
\end{align}
We first show that the constrained gradient of the second through fourth term in \eqref{eq:contastIVE} is equal to zero, so only the derivative of the first term is needed. To this end, note that the derivative of the second term is equal to -$\sum_{k=1}^K{\bf a}_{k,t}$, which follows from \eqref{eq:deriv_identity1b}. It can be shown that the same value, just with a negative sign, is obtained when taking the derivation of the third and fourth terms\footnote{This follows from Equation 33 and Appendix~C in \cite{koldovsky2019TSP}, where the computations were considered for the case $K=1$ and $T=1$.}. Hence, the second, the third and the fourth terms cancel each other out.


Consequently, $\Delta_{k}$ is equal to the derivative of the first term, which, using \eqref{eq:deriv_identity1a}, results in
\begin{equation}\label{eq:diff_bogivew}
\Delta_{k}= 
\frac{1}{T}\sum_{t=1}^T \left\{{\bf \Re}(\nu_{k,t}){\bf a}_{k,t} - \hat{\rm E}\left[\phi_k\left(\left\{\frac{\hat{ s}_{k,t}}{\hat\sigma_{k,t}}\right\}_k\right) \frac{{\bf x}_{k,t}}{\hat\sigma_{k,t}}\right] \right\},
\end{equation}
where ${\bf \Re}(\nu_{k,t})$ denotes the real part of $\nu_{k,t}$, 
\begin{equation}\label{eq:nu_bogivew}
    \nu_{k,t} = \hat{\rm E}\left[\phi_k\left(\left\{\frac{\hat{ s}_{k,t}}{\hat\sigma_{k,t}}\right\}_k\right) \frac{\hat{ s}_{k,t}}{\hat\sigma_{k,t}}\right],
\end{equation}
and $\phi_k(\cdot) = -\frac{\partial}{\partial s_k} \log f(\cdot)$ is the score function corresponding to the model pdf $f(\cdot)$ with respect to the $k$th variable.

To verify the consistency, consider a case when $N \rightarrow +\infty$ and ${\bf w}_k$ is the true separating vector. Then, ${\bf w}_k^H{\bf x}_{k,t}=s_{k,t}$ and since ${\bf x}_{k,t}={\bf a}_{k,t}s_{k,t}+{\bf y}_{k,t}$ and $s_{k,t}$ and ${\bf y}_{k,t}$ are independent, \eqref{eq:diff_bogivew} simplifies to
\begin{equation}
   \Delta_{k} = \frac{1}{T}\sum_{t=1}^T({\bf \Re}(\nu_{k,t}) - \nu_{k,t}){\bf a}_{k,t}.
\end{equation}
It follows that the true separating vector is a stationary point (the gradient is zero) only if ${\bf \Re}(\nu_{k,t}) = \nu_{k,t}$. However, this equality does not hold in general. To overcome this problem, $\phi_k(\cdot)$ can be replaced by $\nu_{k,t}^{-1}\phi_k(\cdot)$ \cite{eusipco2017}. The replacement is possible as this change is equivalent to modifying the model pdf $f(\cdot)$. 
After this modification, the gradient reads
\begin{equation}\label{eq:delta_bogivew}
    \Delta_{k} = \frac{1}{T}\sum_{t=1}^T\left\{{\bf a}_{k,t} - \nu_{k,t}^{-1}\hat{\rm E}\left[\phi_k\left(\left\{\frac{\hat{ s}_{k,t}}{\hat\sigma_{k,t}}\right\}_k\right) \frac{{\bf x}_{k,t}}{\hat\sigma_{k,t}}\right]\right\},
\end{equation}
which is zero when $N \rightarrow +\infty$ and ${\bf w}_k$ is the true separating vector.

Finally, the update rule of BOGIVE$_{\bf w}$ is
\begin{equation}\label{eq:updatew}
    {\bf w}_k \leftarrow {\bf w}_k + \mu \Delta_{k},
\end{equation}
where $\mu>0$ is a step size parameter.
A detailed summary of BOGIVE$_{\bf w}$ is given in Algorithm~\ref{algorithm:BOGIVEw}, in which  the method is started from the initial values of the separating vectors. 
After each iteration, the separating vectors are normalized so that their first elements are equal to $1$ to stabilize the convergence. Once the convergence is achieved, the separating vectors are re-scaled using least squares to reconstruct the images of the SOI on a reference microphone \cite{koldovsky2017}, which is the typical solution to the scaling ambiguity problem.

\begin{algorithm}
\caption{BOGIVE$_{\bf w}$: Block-wise orthogonally constrained independent vector extraction\label{algorithm:BOGIVEw}}
\SetAlgoLined
\KwIn{$\x_{k,t},{\bf w}^{\rm ini}_k$ ($k,t=1,2,\dots$), $\mu$, ${\tt tol}$}
\KwOut{${\bf a}_{k,t},{\bf w}_{k}$}
\ForEach{$k=1,\dots,K$, $t=1,\dots,T$}{
	$\wCxkt=\hat{\rm E}[\x_{k,t}\x_{k,t}^H]$;\\
	${\bf w}_k={\bf w}^{\rm ini}_k/({\bf w}^{\rm ini}_k)_1$;}
\Repeat{$\max\{\|\Delta_{1}\|,\dots,\|\Delta_{K}\|\}<{\tt tol}$}{
	\ForEach{$k=1,\dots,K$, $t=1,\dots,T$}{
		${\bf a}_{k,t} \leftarrow ({\bf w}_k^H\wCxkt{\bf 
			w}_k)^{-1}(\wCxkt{\bf w}_k)$;\\
		$s_{k,t} \leftarrow {\bf w}_{k}^H\x_{k,t}$\\
			$\hat\sigma_{k,t}  \leftarrow \sqrt{{\bf w}_k^H\wCxkt{\bf 
			w}_k}$;}
	\ForEach{$k=1,\dots,K$, $t=1,\dots,T$}{
		Compute $\nu_{k,t}$ according to \eqref{eq:nu_bogivew};}
	\ForEach{$k=1,\dots,K$}{	
	    Compute $\Delta_{k}$ according to \eqref{eq:delta_bogivew};\\
		${\bf w}_k \leftarrow {\bf w}_k + \mu\Delta_{k}$;\\
		${\bf w}_k \leftarrow {\bf w}_k/({\bf w}_k)_1$;}
}
\end{algorithm}



\subsection{Auxiliary Function-Based Algorithm}\label{sec:blockAuxIVE}
In \cite{ono2011stable}, N.~Ono derived the AuxIVA algorithm using an auxiliary function-based optimization (AFO) technique. This method provides a much faster and more stable alternative to the natural gradient-based algorithm from \cite{kim2007}. The main principle of the AFO technique lies in replacing the first term in \eqref{eq:contastIVE} by a majorizing term involving an auxiliary variable that is easier to optimize. The modified contrast function, named the auxiliary function, has the same global maximum as the original contrast function. The auxiliary function is optimized in the auxiliary and normal variables alternately. 

Very recently, a modification of AuxIVA for the blind extraction of $m$ sources, where $m<d$, has been proposed in \cite{scheibler2019}; the algorithm is named OverIVA. In this section, we will apply the AFO technique to find the maximum of \eqref{eq:contastIVE}. The resulting algorithm, which could be seen as a special variant of OverIVA designed for $m=1$ and extended for $T>1$, will be called Block AuxIVE.

Following the same assumption about the model density $f(\cdot)$ as in Theorem~1 in \cite{ono2011stable}, the auxiliary function for \eqref{eq:contastIVE} can have the form
\begin{multline}\label{eq:auxiliaryfunctionCSV}
    Q\left(\{{\bf w}_k,{\bf a}_{k,t},{\bf V}_{k,t}\}_{k,t}\right)= \frac{1}{T}\sum_{t = 1}^{T}\Bigg\{\sum_{k=1}^K -\frac{1}{2} \frac{{\bf w}_k^H {\bf V}_{k,t}{\bf w}_k}{\hat{\sigma}_{k,t}^2} \\-\log \hat{\sigma}_{k,t}^2- \hat{\rm E}[\hat{\bf z}_{k,t}^H{\bf C}_{{\bf z}_{k,t}}^{-1}\hat{\bf z}_{k,t}] +(d-2)\log |\gamma_{k,t}|^2 \Bigg\} + R_t,  
\end{multline}
 where 
\begin{equation}\label{eq:Vkt}
    {\bf V}_{k,t} = \hat{\rm E}[\varphi(r_t){\bf x}_{k,t}{\bf x}_{k,t}^H]
\end{equation}
are the auxiliary variables, and $R_t$ depends purely on $r_t$; $\varphi(\cdot)$ is an appropriate nonlinearity chosen such that the conditions of Theorem~1 in \cite{ono2011stable} are satisfied; its concrete choice will be discussed in Section~\ref{sec:nonlinearity}. Now, it holds that 
\begin{equation}\label{eq:auxiliaryEquality}
    \mathcal{C}(\{{\bf w}_k,{\bf a}_{k,t}\}_{k,t}) \leq Q(\{{\bf w}_k,{\bf a}_{k,t},{\bf V}_{k,t}\}_{k,t}),
\end{equation}
where both sides are equal when $r_t = \sqrt{\sum_{k=1}^K |{\bf w}_k^H{\bf x}_{k,t}|^2}$ for every $t=1,\dots,T$.

In a way similar to Section~\ref{sec:OGIVEw}, the OGC is imposed between the pairs of vector variables ${\bf w}_k$ and ${\bf a}_{k,t}$. The optimization of $ Q$ proceeds alternately in the auxiliary variables ${\bf V}_{k,t}$ and the normal variables ${\bf w}_k$. The maximum of \eqref{eq:auxiliaryfunctionCSV} in the auxiliary variables is obtained by putting $r_t = \sqrt{\sum_{k=1}^K |{\bf w}_k^H{\bf x}_{k,t}|^2}$ into \eqref{eq:Vkt}. To find the minimum in the normal variables, the constrained gradients of \eqref{eq:auxiliaryfunctionCSV} are set to zero. Since the second, third and fourth term of \eqref{eq:auxiliaryfunctionCSV} are equal to those in \eqref{eq:contastIVE}, only the first term in \eqref{eq:auxiliaryfunctionCSV} contributes to the gradient, which is
\begin{multline}
\label{eq:blockauxivederivace}
\left.\frac{\partial  Q\left(\{{\bf w}_k,{\bf V}_{k,t}\}_{k,t}\right)}{\partial {\bf w}_k^H}\right|_\text{w.r.t. \eqref{eq:couplingaCSV}}= \\
\frac{1}{T}\sum_{t=1}^{T}\left\{\frac{{\bf w}_k^H {\bf V}_{k,t}{\bf w}_k}{\hat{\sigma}_{k,t}^2}{\bf a}_{k,t}-\frac{{\bf V}_{k,t}{\bf w}_k}{\hat{\sigma}_{k,t}^2}\right\}.
\end{multline}
The close-form solution of the equation when \eqref{eq:blockauxivederivace} is put equal to zero cannot be derived in general. Our proposal is to take 
$
\w =  \left(\sum_{t=1}^{T}\frac{\V}{\hat{\sigma}_{k,t}^2}\right)^{-1} \sum_{t=1}^{T}\frac{{\bf w}_k^H {\bf V}_{k,t}{\bf w}_k}{\hat{\sigma}_{k,t}^2}{\bf a}_{k,t},
$
which is the solution of a linearized equation where the terms ${\bf w}_k^H {\bf V}_{k,t}{\bf w}_k$ and $\hat{\sigma}_{k,t}^2$ are treated as constants that are independent of ${\bf w}_k$. 
Hence, the update rules of Block AuxIVE are as follows:
\begin{align}
    r_t &=\sqrt{\sum_{k=1}^K |{\bf w}_k^H{\bf x}_{k,t}|^2}\label{eq:blockauxiverule1}, \\
    \V &= \hat{\rm E}\left[\varphi(r_t){\bf x}_{k,t}{\bf x}_{k,t}^H\right],\\
{\bf a}_{k,t} & =\frac{\wCxkt{\bf w}_k}{{\bf w}_k^H\wCxkt{\bf w}_k},\\
\sigma_{k,t} &= \sqrt{{\bf w}_k^H\wCxkt{\bf w}_k}\\
\w &=  \left(\sum_{t=1}^{T}\frac{\V}{\hat{\sigma}_{k,t}^2}\right)^{-1} \sum_{t=1}^{T}\frac{{\bf w}_k^H {\bf V}_{k,t}{\bf w}_k}{\hat{\sigma}_{k,t}^2}{\bf a}_{k,t}\label{eq:blockauxiverule5},\\
\w &\leftarrow \w/\sqrt{\sum_{t=1}^{T}{\bf w}_k^H {\bf V}_{k,t}{\bf w}_k}.
\end{align}
The last step, which performs a normalization of the updated separating vectors, has been found important to the stability of the convergence. After the convergence is achieved, the scaling ambiguity problem is resolved in the same way as in BOGIVE$_{\bf w}$.
The pseudo-code is summarized in Algorithm~\ref{alg:Auxive}. 

\begin{algorithm}
    \caption{Block AuxIVE\label{alg:Auxive}}
    \SetAlgoLined
    \KwIn{$\x_{k,t},{\bf w}^{\rm ini}_k$ ($k,t=1,2,\dots$), ${\tt NumIter}$}
\KwOut{${\bf a}_{k,t},{\bf w}_{k}$}
\ForEach{$k=1,\dots,K$, $t=1,\dots,T$}{
	$\wCxkt=\hat{\rm E}[\x_{k,t}\x_{k,t}^H]$;\\
	${\bf w}_k={\bf w}^{\rm ini}_k/({\bf w}^{\rm ini}_k)_1$;}
	${\tt Iter} = 0$;\\
\Repeat{${\tt Iter}<{\tt NumIter}$}
    {
    \ForEach{$t = 1 \dots T$}
        {
        {$r_t \leftarrow\sqrt{\sum_{k = 1}^{K}|\w^H{\bf x}_{k,t}|^2}$};\\
        \ForEach{$k = 1 \dots K$}
        {
            ${\bf a}_{k,t} \leftarrow ({\bf w}_k^H\wCxkt{\bf 
			w}_k)^{-1}(\wCxkt{\bf w}_k)$;\\
            $\hat\sigma_{k,t} \leftarrow {\bf w}_k^H\wCxkt{\bf w}_k $
           $\V  \leftarrow \hat{\rm E}[\varphi(r_t){\bf x}_{k,t}{\bf x}_{k,t}^H]$;\\
        }
        }
         \ForEach{$k = 1 \dots K$}
        {
        Compute $\w^H$ according \eqref{eq:blockauxiverule5};\\
        ${\bf w}_{k} \leftarrow {\bf w}_k/\sqrt{\sum_{t=1}^{T}{\bf w}_k^H {\bf V}_{k,t}{\bf w}_k}$;\\
        }
     ${\tt Iter} \leftarrow {\tt Iter} +1$;\\       
    }
\end{algorithm}




\subsection{Piloted Block AuxIVE}\label{sec:piloted}

Owing to the indeterminacy of BSE it is not, in general, known which source is currently being extracted. The crucial problem is to ensure that the signal being extracted actually corresponds to the desired SOI. In BOGIVE$_{\bf w}$ as well as in Block AuxIVE, this can be influenced only through the initialization.

Several approaches ensuring the global convergence have been proposed, most of which are based on additional constraints assuming prior knowledge, e.g., about the source position or a reference signal \cite{parra2002,khan2015,bhinge2019,brendel2020}. Recently, an unconstrained supervised IVA using the so-called pilot signals has been proposed in  \cite{nesta2017supervised}, where each pilot signal is dependent on the source signals, which \redalert{means} that they have a joint pdf that cannot be factorized into a product of marginal pdfs. This idea has been extended to IVE in \cite{koldovsky2019TSP}, where the pilot signal related to the SOI is assumed to be available. 

Let the pilot signal (dependent on the SOI and independent of the background) be represented, on the $t$th block, by $\pilot_t$ ($\pilot_t$ is denoted without index $k$, nevertheless, it can also be $k$-dependent).  Let the joint pdf of ${\bf s}_t$ and $o_t$ be $p({\bf s}_t,\pilot_t)$. Then, similarly to \eqref{eq:independence2}, the pdf of the observed data within the $t$th block is given by 
\begin{multline}
 p_{\bf x}(\{{\bf x}_{k}\}_{k,t}) = p(\{{\bf w}_k^H{\bf x}_{k,t}\}_{k,t},\pilot_t)\times\\ \prod_{k=1}^K p_{{\bf z}_{k,t}}({\bf B}_{k,t}{\bf x}_{k,t}) |\det {\bf W}_{k,t}|^2.
\end{multline}
Comparing this expression with \eqref{eq:independence2} and taking into account the fact that $\pilot_t$ is independent of the mixing model parameters, it can be seen that the modification to use pilot signals within Block AuxIVE is straightforward. 

In particular, provided that the model pdf $f(\{{\bf w}_{k}^H {\bf x}_k\}_{k,t},\pilot_t)$ replacing the unknown $p(\cdot)$ meets the conditions for the application of AFO as in Section~\ref{sec:blockAuxIVE}, the piloted algorithm has exactly the same steps as the non-piloted one with a sole difference that the non-linearity $\varphi(\cdot)$ also depends on $\pilot_t$. 
The equality between the contrast function and the auxiliary function 
holds if and only if
\begin{equation}\label{eq:auxivepiloted}
r_t = \sqrt{\sum_{k=1}^K |({\bf w}_k)^H{\bf x}_{k,t}|^2 + \cpilot^2|\pilot_t|^2},
\end{equation}  
for $t=1,\dots,T$, where $\cpilot$ is a hyperparameter controlling the influence of the pilot signal \cite{nesta2017supervised}. 
Finally, Piloted Block AuxIVE is obtained from Block AuxIVE by replacing the update step \eqref{eq:blockauxiverule1} with \eqref{eq:auxivepiloted}. 

Finding a suitable pilot signal poses an application-dependent problem. For example, outputs of voice activity detectors were used to pilot the separation of simultaneously talking people in \cite{nesta2017supervised}. Similarly, a video-based lip-movement detection was considered in \cite{nesta2017eusipco}. A video-independent solution was proposed in \cite{cmejla2018} using spatial information about the area in which the speaker is located. Recently, the approach utilizing speaker identification by neural networks was proposed in \cite{Jansky2020}. All these approaches have been shown useful, although the pilot signals used contain residual noise and interference. The choice of a piloting signal for this paper will be discussed in Section IV.

\subsection{Choice of $f(\cdot)$}\label{sec:nonlinearity}

In this paper, we choose the model pdf in the same way as it was proposed in the pioneering IVA paper \cite{kim2007}; namely,
\begin{equation}
   f({\bf s}_t)\propto \exp\{-\|{\bf s}_t\|\},
\end{equation}
for which the $k$th score function is
\begin{equation}
    \phi_k({\bf s}_t)=-\frac{\partial}{\partial s_k} \log f({\bf s}_t)=\frac{s_{k,t}}{\|{\bf s}_t\|},
\end{equation}
and the related nonlinearity in \eqref{eq:Vkt} is $\varphi(\|{\bf s}_t\|)=\|{\bf s}_t\|^{-1}$. This pdf satisfies the conditions for applying AFO (Theorem~1 in \cite{ono2011stable}) and is known to be suitable for speech signals that are typically super-Gaussian.
It is worth noting here that  more accurate modeling of the source  pdf usually leads to improved performance. For example, advanced statistical models are currently studied for ILRMA \cite{kitamura2018,mogami2019}. However, this topic goes beyond the scope of this work.


\section{Experimental Validation}\label{sec:experiments}
In this section, we present results of experiments with simulated mixtures as well as real-world recordings of moving speakers. Our goal is to show the usefulness of the CSV mixing model and compare the performance characteristics of the proposed algorithms with other methods assuming the conventional static mixing models.

\subsection{Simulated room} \label{sec:experiment_simulated}
In this example, we inspect de-mixing filters obtained by the blind algorithms when extracting a moving speaker in a room simulated by the image method \cite{allen1979}. The room has dimensions $4\times4\times2.5$ (width$\times$length$\times$height) metres and $T_{60}= 100$~ms.  A linear array of five omnidirectional microphones is located so that its center is at the position $(1.8, 2, 1)$~m, and the array axis is parallel with the room width. The spacing between microphones is $5$~cm. 

\begin{figure*}[t]
\begin{subfigure}[t]{0.32\linewidth}
    \includegraphics[width=1\textwidth]{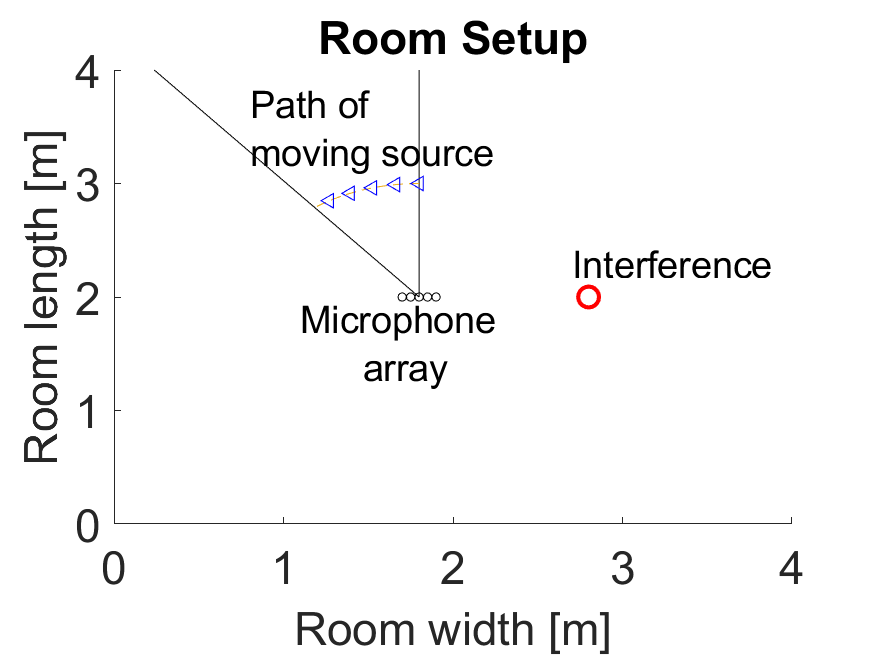}
      \caption{Setup of the simulated room conditions. The position of interference is marked by the red circle, the microphones by black circles and the path of the source is marked by the blue triangles.}
      \label{fig:room_setup}
\end{subfigure}
\hspace{0.015\linewidth}
\begin{subfigure}[t]{0.32\linewidth}
 \includegraphics[width=1\textwidth]{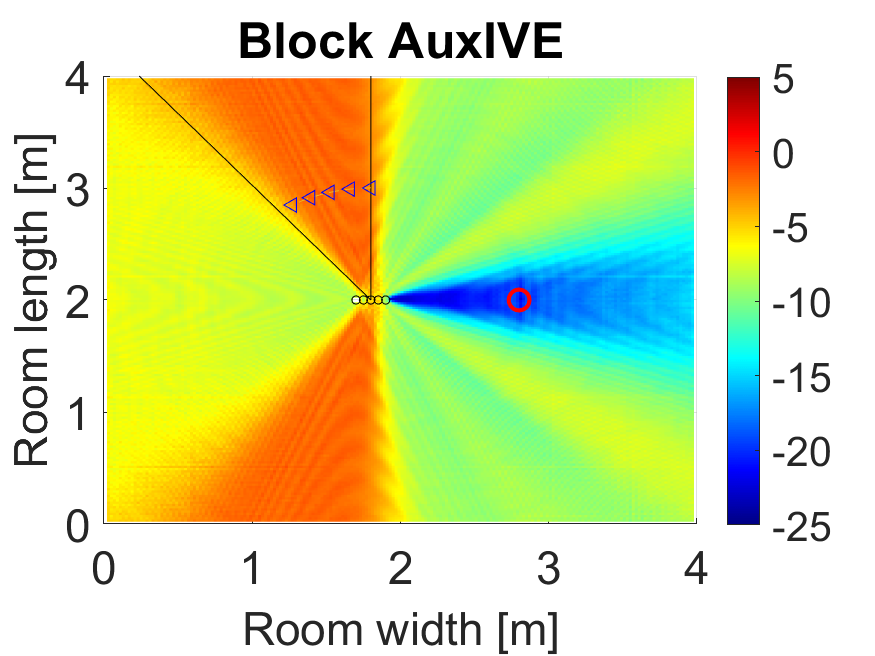}
 \caption{Attenuation in dB achieved by Block AuxIVE}
  \label{fig:auxbogive}
\end{subfigure}
\hspace{0.015\linewidth}
\begin{subfigure}[t]{0.32\linewidth}
 \includegraphics[width=1\textwidth]{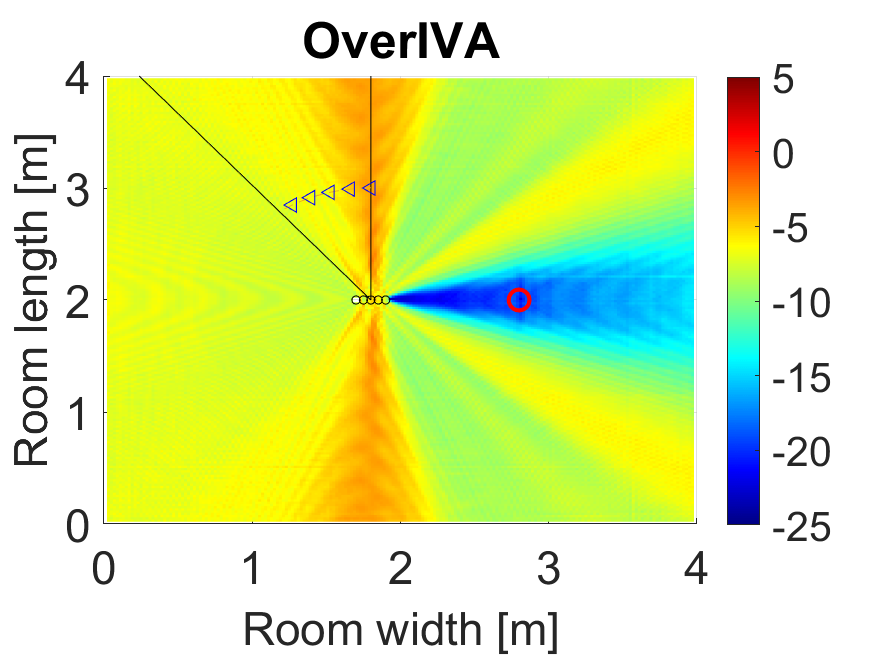}
 \caption{Attenuation in dB achieved by OverIVA}

  \label{fig:auxogive}
\end{subfigure}
\vspace*{0.2cm}

\begin{subfigure}[t]{0.32\linewidth}
 \includegraphics[width=1\textwidth]{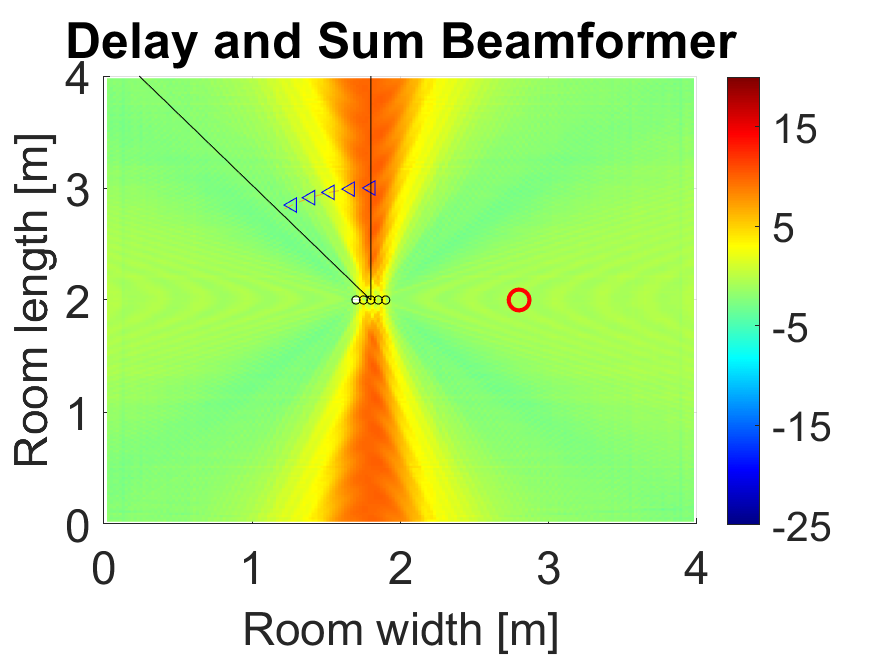}
 \caption{Attenuation in dB achieved by Delay and sum Beamformer}
  \label{fig:doa}
\end{subfigure}
\hspace{0.015\linewidth}
\begin{subfigure}[t]{.32\textwidth}
 \includegraphics[width=1\textwidth]{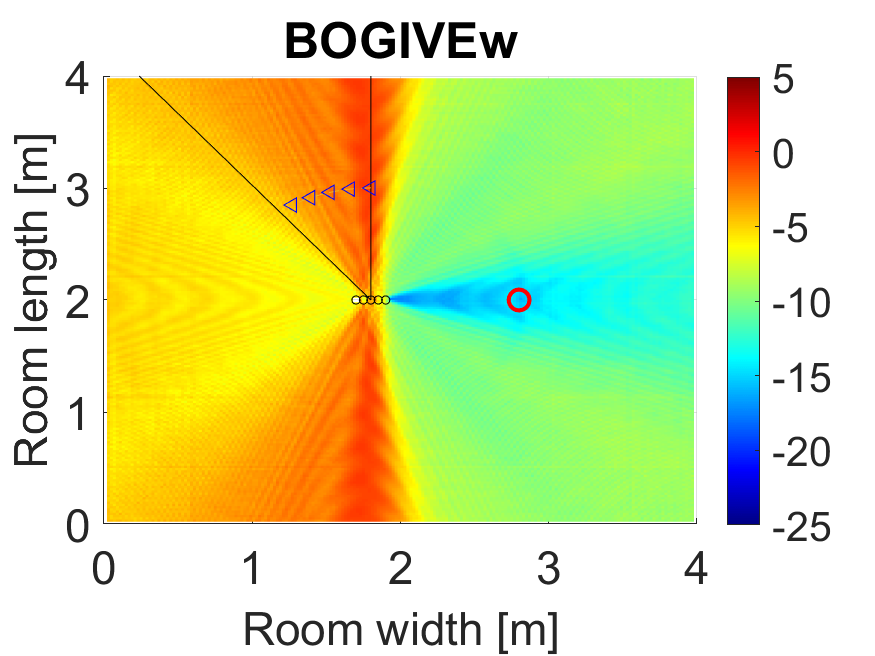}
 \caption{Attenuation in dB achieved by $\text{BOGIVE}_\text{w}$ }
  \label{fig:bogive}
\end{subfigure}
\hspace{0.015\linewidth}
\begin{subfigure}[t]{0.32\linewidth}
\includegraphics[width=1\textwidth]{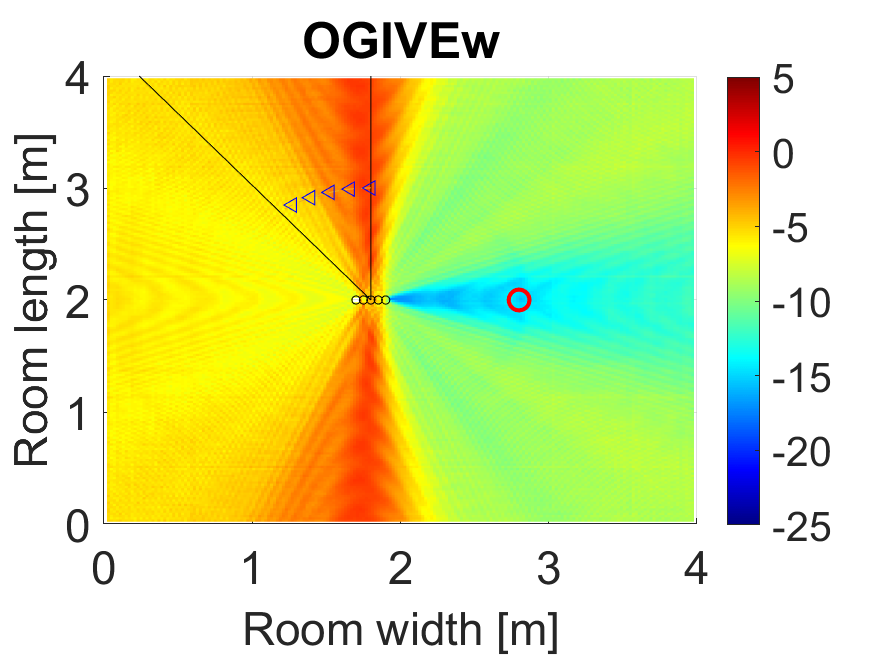}
 \caption{Attenuation in dB achieved by $\text{OGIVE}_\text{w}$}
  \label{fig:ogive}
\end{subfigure}
\caption{Setup of the simulated room and the attenuation in dB achieved by DOA, {$\text{OGIVE}_\text{w}$}, {$\text{BOGIVE}_\text{w}$}, {OverIVA} and {Block AuxIVE} from the experiment in Section~\ref{sec:experiment_simulated}}

\end{figure*}

The target signal is a 10~s long female utterance from TIMIT. During speech, 
the speaker is moving at a constant speed on a $38^{\circ}$ arc at a one-meter 
distance from the center of the array; the situation is illustrated in 
Fig.~\ref{fig:room_setup}. The starting and ending positions are $(1.8, 3, 
1)$~m and  {$(1.2,2.78,1)$}~m, respectively. The movement is simulated by 20 
equidistantly spaced RIRs on the path, which correspond to half-second 
intervals of speech, 
whose overlap was smoothed by windowing. Next, a point source emitting a white Gaussian noise is located at the position $(2.8, 2, 1)$~m; that is, at a one-meter distance to the right from the array. 

The mixture of speech and noise has been processed in order to extract the speech signal by the following methods: {$\text{OGIVE}_\text{w}$}, {$\text{BOGIVE}_\text{w}$}, {OverIVA} with $m=1$ \cite{scheibler2019}, which corresponds with Block AuxIVE when $T=1$, and {Block AuxIVE}. All methods operate in the STFT domain with the FFT length of $512$ samples and $128$ samples hop-size; the sampling frequency is $f_s=16$ kHz. Each method has been initialized by the direction of arrival of the desired speaker signal at the beginning of the sequence. The other parameters of the methods are listed in Table~\ref{tab:exp1setup}.  
\begin{table}
\caption{Parameter setup for the tested methods in the simulated room}
\label{tab:exp1setup}
\centering
\begin{tabular}{l|ccc}
Method                        & \multicolumn{1}{l}{\# iterations} & \multicolumn{1}{l}{step size $\mu$} & \multicolumn{1}{l}{block size $N_b$} \\\hline
$\text{OGIVE}_\text{w}$       & 1000                                     & 0.2                                    & n/a                              \\
$\text{BOGIVE}_\text{w}$ & 1000                                     & 0.2                                    & 250 frames                     \\
OverIVA                        & 100                                      & n/a                                      & n/a                              \\
Block AuxIVE                  & 100                                      & n/a                                       & 250 frames                    
\end{tabular}
\end{table}

In order to visualize the performance of the extracting filters, a $2\times2$~cm-spaced regular grid of positions spanning the whole room is considered.  Microphone responses (images) of a  white Gaussian noise signal emitted from each position on the grid have been simulated. The extracting filter of a given algorithm is applied to the responses, and the output power is measured. The average ratio between the output power and the power of the input signals reflects the attenuation of the white noise signal originating from the given position. 
 
The attenuation maps of the compared methods are shown in Figures \ref{fig:auxbogive} through \ref{fig:ogive}, and Table \ref{tab:att_results} shows the attenuation for specific points in the room. In particular, the first five columns in the table correspond to the speaker's positions on the movement path at angles $0^{\circ}$ through $32^{\circ}$. The last column corresponds to the position of the interferer. 

Fig.~\ref{fig:doa} shows the map of the initial filter corresponding to the delay-and-sum (D\&S) beamformer steered towards the initial position of the speaker. The beamformer yields a gentle gain in the initial direction with no attenuation in the direction of the interferer.

By contrast, all the compared blind methods steer a spatial null towards the interferer and try to increase the gain of the target signal. The spatial beam steered by Block AuxIVE towards the speaker spans the whole angular range where the speaker has appeared during the movement. $\text{BOGIVE}_\text{w}$ performs similarly. However, its performance is poorer, perhaps due to its slower convergence or proneness to getting stuck in a local extreme. OverIVA and $\text{OGIVE}_\text{w}$ tend to focus on only a narrow angular range (probably the most significant part of the speech). The nulls steered towards the interferer by OverIVA and Block AuxIVE are more attenuating compared to the gradient methods. 
In conclusion, these results confirm the validity of the CSV mixing model and show the better convergence properties of Block AuxIVE over $\text{BOGIVE}_\text{w}$.

\begin{table}
\caption{The attenuation (dB) in selected points on the source path and in the position of the interferer \label{tab:att_results}}
\begin{tabular}{lllllll}
{}      & $0^{\circ}$            & $8^{\circ}$             & $16^{\circ}$            & $24^{\circ}$            & $32^{\circ}$           & Interferer   \\ 
\hline
$\text{OGIVE}_\text{w}$       & \textbf{-1.09} & \textbf{-1.36} & -2.02  & -4.56   & -5.08  & -15.81         \\ 
$\text{BOGIVE}_\text{w}$ & -1.20          & -2.14         & -1.69  & -3.12   & -3.87 & -15.86          \\
OverIVA    & -5.85  & -3.99    & -3.08  & -4.39   & -5.12 & \textbf{-23.73} \\
Block AuxIVE  & -3.22  & -1.74   & \textbf{-1.27} & \textbf{-2.09} & \textbf{-2.67} & -18.51         
\end{tabular}
\end{table}

 \begin{figure*}
     \includegraphics[width=1\textwidth]{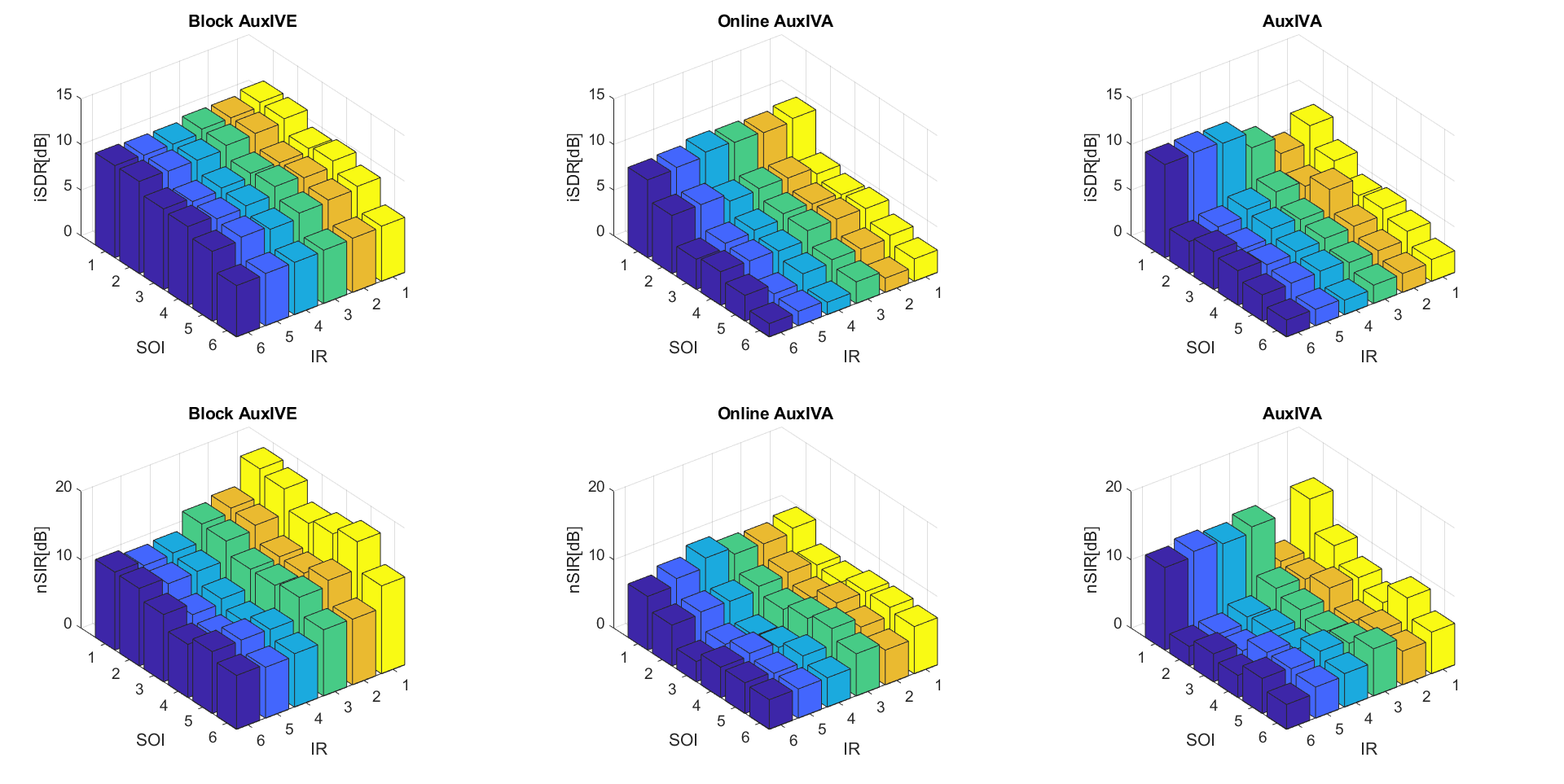}
      \includegraphics[width=1\textwidth]{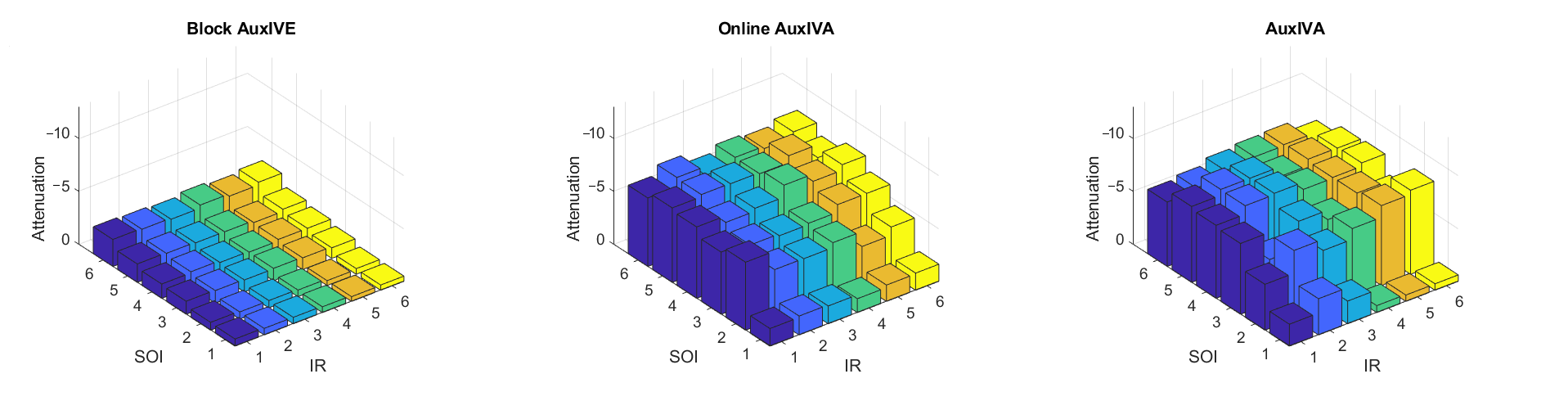}
  \caption{\redalert{The accuracy of the blind extraction of the SOI in terms of iSDR, nSIR and Attenuation in the experiment in Section~\ref{sec:loudspeakerexperiment}. The indices on the SOI and IR axes correspond with Table~\ref{tab:exp1setup}. Please note that, for better readability, the axes of the plots showing the Attenuation are {\em reversed}.}}
     \label{fig:realBar}
 \end{figure*}

{

\subsection{Moving speakers simulated by wireless loudspeaker attached to turning arm}\label{sec:loudspeakerexperiment}

The goal of this experiment is to compare the performance of algorithms as they depend on the range and speed of movements of sources. We have recorded a dataset of speech utterances that were played from a wireless loudspeaker (JBL GO 2) attached to a manually actuated rotating arm. The length of each utterance is $31$~s. Sounds were recorded using a linear array of four microphones with $16$~cm spacing. The array center was placed at the arm's pivot. This allows the apparatus to simulate circular movements of sources at a radius of $\approx1$~m. The recording setup was placed in an open-space $12$x$8$x$2.6$~m room with a reverberation time $\mathrm{T_{60}}\approx 500 \mathrm{ms}$.



The dataset consists of two individual, spatially separated sources. The SOI is represented by a male speech utterance and is confined to the angular interval from 0\degree through 90\degree. The interference (IR) is represented by a female speech utterance and is confined to the interval of -90\degree through 0\degree. The list of recordings is described in Table~\ref{tab:Real_setup}. The recordings along with videos of the recording process are available online\footnote{\tt \url{https://asap.ite.tul.cz/downloads/ice/blind-extraction-of-a-moving-speaker/ %https://drive.google.com/drive/folders/1rBlc3yuPhNO-AWibHpu8MfGtxOAFBVlN?usp=sharing
}}.

\begin{table*}[]
\caption{\redalert{Angular intervals and speed of SOI and IR movements in experiment \ref{sec:loudspeakerexperiment}.}} 
\label{tab:Real_setup}
\centering
\redalert{
\begin{tabular}{l|c|c|c|c|c|c}
Recording index & 1& 2  & 3 & 4 & 5  & 6 \\ \hline
Movement speed & static  & slow & fast & slow & fast & very fast \\ \hline
\multicolumn{7}{c}{Range of the movement: starting angle, ending angle} \\ \hline
SOI (male) & \multicolumn{1}{r|}{$-45$\degree} & \multicolumn{1}{r|}{ $-55$\degree, $-35$\degree} & \multicolumn{1}{r|}{$-55$\degree, $-35$\degree} & \multicolumn{1}{r|}{ $-90$\degree ,$0$\degree } & \multicolumn{1}{r|}{ $-90$\degree, $0$\degree} & \multicolumn{1}{r}{$-90$\degree, $0$\degree} \\
IR (female) & \multicolumn{1}{r|}{$45$\degree}  & \multicolumn{1}{r|}{$35$\degree, $55$\degree }    & \multicolumn{1}{r|}{ $35$\degree, $55$\degree}  & \multicolumn{1}{r|}{ $0$\degree, $90$\degree }   & \multicolumn{1}{r|}{ $0$\degree, $90$\degree }   & \multicolumn{1}{r}{$0$\degree, $90$\degree}  
\end{tabular}}
\end{table*}

$36$ mixtures were created by combining the SOI and IR recordings in Table~\ref{tab:Real_setup}; the input SIR was set to $10$~dB. The following three algorithms were compared: Block AuxIVE with the length of blocks set to $100$ frames, the original AuxIVA algorithm \cite{ono2011stable}, and a sequential on-line variant of AuxIVA (On-line AuxIVA) from \cite{taniguchi2014} with the time-window length of $20$ frames and the forgetting factor set to $0.95$. The algorithms operated in the STFT domain with $1024$ samples per frame and $768$ samples overlap. The off-line algorithms were stopped after $100$ iterations. 

Performance was evaluated using segmental measures: normalized SIR (nSIR), SDR improvement (iSDR), and the average SOI attenuation (Attenuation); nSIR is the ratio of the powers of the SOI and IR in the extracted signal whose each segment is normalized to unit variance; SDR is computed using BSS\_eval \cite{bsseval}. While iSDR and Attenuation reflect the loss of power of the SOI in the extracted signal, nSIR reflects also the IR cancellation. The length of segments was set to $1$~s.

The results are shown in Fig.~\ref{fig:realBar}. AuxIVA and On-line AuxIVA achieve their optimum performances in terms of the criteria when the SOI is static (SOI index 1). Their performances drop when the SOI moves (SOI index $>1$). On-line AuxIVA is slightly less sensitive to the SOI movements compared to AuxIVA due to its adaptability. However, the overall performance of On-line AuxIVA is low, because the algorithm performs locally and cannot exploit a larger batch of data to achieve higher extraction accuracy.

Block AuxIVE shows significantly smaller sensitivity to the SOI movements than the compared algorithms. This is mainly reflected by Attenuation, which is only slightly growing with the increasing range and speed of the SOI movements (SOI index $>1$). The superior performance of Block AuxIVE in terms of iSDR and nSIR compared to AuxIVA and On-line AuxIVA demonstrates the new quality of the proposed algorithm gained due to the CSV mixing model.

The IR movements cause the performance of AuxIVA and Block AuxIVE to decrease with the growing range of the IR movement (IR index $>1$). The speed of movement seems to play a minor role. This can be explained as follows. The off-line algorithm should ideally estimate a spatial filter which projects two distinct wide beams: one towards the entire angular area occupied by the SOI and one towards the area occupied by the IR. The former beam should pass the incoming signal through while the latter beam should attenuate it. Provided that the estimated filter satisfies these requirements, as long as the sources stay within their respective beams, the speed with which they move shouldn't matter.

In conclusion, the results reflect the theoretical capabilities of the algorithms, or, more specifically, of the filters that they can estimate. AuxIVA can steer only a narrow beam towards the SOI, which can therefore be extracted  efficiently only if the SOI is not moving. 
Also, On-line AuxIVA can steer narrow beams only, nevertheless, it can adapt them towards the actual positions of the sources at a specific time. This, however, requires precise tracking of the sources, for which a sufficient context of data is needed. Once the sources move too fast, they can no longer be described as point sources within a longer context; a compromise must be made (short enough time-window and low forgetting factor). This is where CSV offers new capabilities. Block AuxIVE based on the CSV model can reliably extract the SOI from a wider area within a context of the data. 
Regarding IR, both AuxIVA and Block AuxIVE can attenuate a moving IR by placing multiple null beams along its path. The number of the null beams, hence the range of the area, is limited by the degrees of freedom, which is given by the number of microphones.
}

\subsection{Real-world scenario using the MIRaGe database}\label{sec:mirageexperiment}

The experiment here is designed to provide an exhaustive test of the compared methods in challenging noisy situations where the target speaker is performing small movements within a confined area. Recordings are simulated using real-world room impulse responses (RIRs) taken from the MIRaGe database \cite{mirage}.

MIRaGe provides measured RIRs between microphones and a source whose possible positions form a dense grid within a $46\times 36\times 32$~cm volume. MIRaGe is thus suitable for our experiment, as it enables us to simulate small speaker movements in a real environment. 

The database setup is situated in an acoustic laboratory which is a $6\times 6\times 2.4$~m rectangular room with variable reverberation time. Three reverberation levels with $\mathrm{T}_{60}$ equal to $100$, $300$, and $600$~ms are provided. The speaker's area involves $4104$ positions which form the cube-shaped grid with spacings of $2$-by-$2$~cm over the $x$ and $y$ axes and $4$~cm over the $z$ axis. Also, MIRaGe contains a complementary set of measurements that provide information about the positions placed around the room perimeter with spacing of $\approx$1~m, at a distance of $1$~m from the wall. These positions are referred to as the out-of-grid positions (OOG). All measurements were recorded by six static linear microphone arrays ($5$ mics per array with the inter-microphone spacing of $-13$, $-5$, $0$, $+5$ and $+13$~cm relative to the central microphone); for more details about the database, see \cite{mirage}. 

In the present experiment, we use Array 1, which is at a distance of $1$~m from the center of the grid, and the T$_{60}$ settings with $100$ and $300$~ms, respectively. For each setting, $3840$ noisy observations of a moving speaker were synthesized as follows: each mixture consists of a moving SOI, one static interfering speaker and noise. The SOI is moving randomly over the grid positions. The movement is simulated so that the position is changed every second. The new position is randomly selected from all positions whose maximum distance from the current position is $4$ in both the $x$ and $y$ axes. The transition between positions is smoothed using the Hamming window of a length of $f_s/16$ with one-half overlaps. The interferer is located in a random OOG position between $13$ through $24$, while the noise signal is equal to a sum of signals that are located in the remaining OOG positions (out of $13$ through $24$).

As the SOI and interferer signal, clean utterances of 4 male and 4 female speakers from  CHiME-4 \cite{chime_data_soft} database were selected; there are $20$ different utterances, each having $10$~s in length per speaker. The noise signals correspond to random parts of the CHiME-4 cafeteria noise recording. The signals are convolved with the RIRs to match the desired positions, and the obtained spatial images of the signals on microphones are summed up so that the interferer/noise power ratio, as well as the power ratio between the SOI and interference plus noise, is $0$~dB.  

\begin{table*}[t]
\caption{The SINR improvement with standard deviation, SDR improvement with standard deviation and extraction fail percentage for the MIRaGe  database experiment}
\label{tab:mirage_results}
\centering
\begin{tabular}{l|ccc|ccc|c}
& \multicolumn{3}{c|}{T60 100 ms} & \multicolumn{3}{c|}{T60 300ms} &\multicolumn{1}{c}{average}\\ \cline{2-7} 
 & \multicolumn{1}{c}{\begin{tabular}[c]{@{}c@{}}mean  iSINR \\ {[}dB{]} \end{tabular}} & \multicolumn{1}{c}{\begin{tabular}[c]{@{}c@{}}mean  iSDR \\ {[}dB{]} \end{tabular}} & \multicolumn{1}{c|}{\begin{tabular}[c]{@{}c@{}} iSINR \textless -5 dB\\ 
 {[}\%{]}\end{tabular}} & \multicolumn{1}{c}{\begin{tabular}[c]{@{}c@{}}mean  iSINR \\{[}dB{]} \end{tabular}}  & \multicolumn{1}{c}{\begin{tabular}[c]{@{}c@{}}mean  iSDR \\ {[}dB{]} \end{tabular}} & \multicolumn{1}{c|}{\begin{tabular}[c]{@{}c@{}}iSINR \textless -5 dB\\  {[}\%{]}\end{tabular}} & \multicolumn{1}{c}{\begin{tabular}[c]{@{}c@{}}time per\\  mixture [s]\end{tabular}}\\ \hline
OverIVA                          &      7.55 $\pm$ 8.33  &      3.96 $\pm$ 2.14  &      8.83 &      5.34 $\pm$ 7.01  &       3.82 $\pm$ 2.00  &      8.43  &  {\bf 8.00}\\
Block AuxIVE                    &      9.45 $\pm$ 7.24  &      4.02 $\pm$ 1.27  &      6.72  &      6.84 $\pm$ 6.52  &       3.48 $\pm$ 1.17  &      6.71  &       9.14 \\
Piloted  OverIVA                 &  7.23 $\pm$ 6.45 &  3.88 $\pm$2.01 &      3.01  &      7.01 $\pm$ 5.22  &   3.55 $\pm$ 1.98 &      2.45   &       8.02\\
Piloted Oracle OverIVA           &      11.99 $\pm$ 5.42  &      5.10 $\pm$ 3.37  &  0.65 &  9.67 $\pm$ 4.58 &       3.00 $\pm$ 2.55  & 0.26  &       8.16\\
Piloted Block AuxIVE            &      8.23 $\pm$ 5.31  &      3.93 $\pm$ 2.78  &  2.53 &  7.52 $\pm$ 4.76 &       3.22 $\pm$ 1.38  &  1.59 &       9.16\\
Piloted Oracle Block AuxIVE           &      {\bf 13.72 $\pm$ 3.51} &      {\bf 6.14 $\pm$ 2.13 } &      {\bf 0} &  {\bf 11.41 $\pm$ 3.54} &       {\bf 4.73 $\pm$ 1.91}  &{\bf  0}  &       9.14\\
$\text{BOGIVE}_\text{w}$   &      4.32 $\pm$ 5.15  &      3.14 $\pm$ 1.56  &      15.32 &      2.28 $\pm$ 3.15  &       1.98 $\pm$ 1.02  &      22.15  &      86.45 \\
$\text{OGIVE}_\text{w}$         &      3.85 $\pm$ 4.33  &      3.58 $\pm$ 1.98  &      22.10 &      1.01 $\pm$ 2.17  &       2.14 $\pm$ 1.45  &      12.23  &      73.15                                                                                           
\end{tabular}
\end{table*}

The methods and their parameters are compared as follows: Additionally to the 
methods compared in the previous section, we consider the piloted variants of   
OverIVA and Block AuxIVA. The number of iterations is set to $150$ and $2,000$ 
for the AFO-based and the gradient-based methods, respectively. The block size 
for the block methods is set to $150$ frames. The gradient step-length for 
{$\text{OGIVE}_\text{w}$} and  {$\text{BOGIVE}_\text{w}$} is set to $\mu = 
0.2$. The initial separating vector $\w$ is initialized by the D\&S beamformer 
steered in front of the microphone array. In the piloted methods, the piloting 
signals are equal to the output of an MPDR beamformer where the steering vector 
corresponds to the ground true DOA of the SOI. { Note that the MPDR output 
provides only a rough estimate of the SOI as the ground true DOA does not take 
the reverberation into account.}  We also consider oracle-piloted variants 
where the pilots corresponds to the ground truth SOI. All these methods operate 
in the STFT domain with an FFT length of $1024$ and a hop-size of $256$; the 
sampling frequency is $16$~kHz.  

\begin{figure*}[t]
 
\begin{subfigure}[t]{.49\textwidth}
 \centering
     \includegraphics[width=1\textwidth]{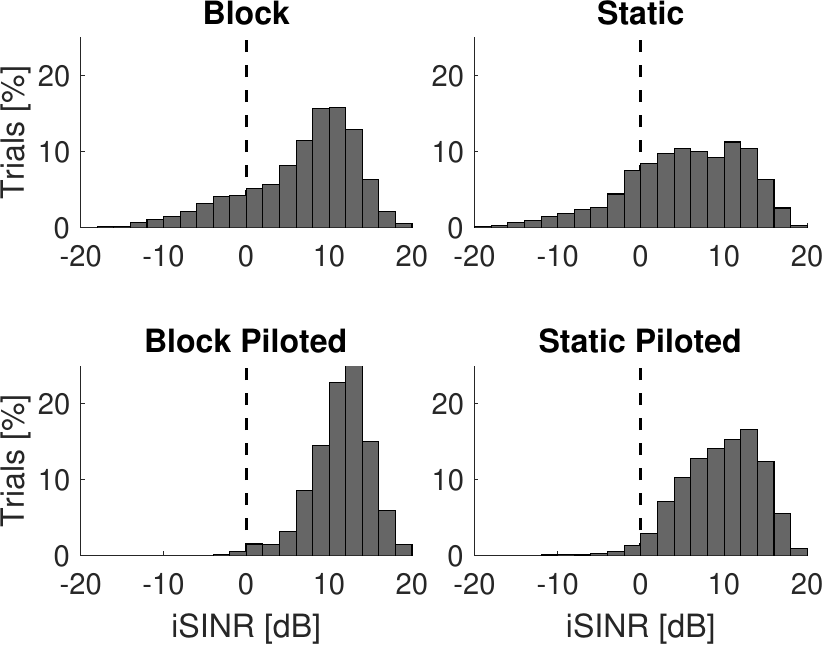}
     \caption{Percentage histogram of SINR improvement over full dataset.}
     \label{fig:hist_full}
\end{subfigure}
\hspace{0.015\linewidth}
    \begin{subfigure}[t]{.49\textwidth}
 \centering
     \includegraphics[width=1\textwidth]{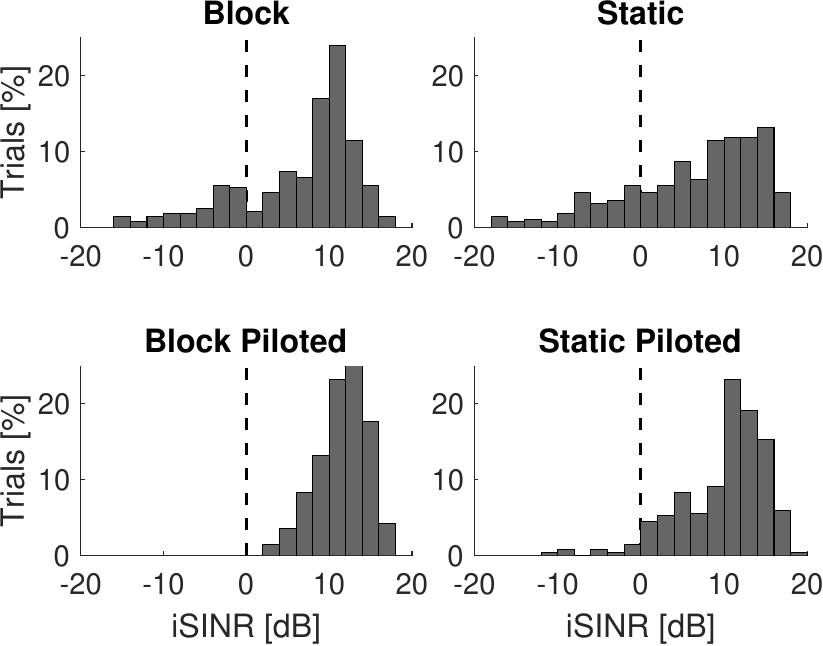}
     \caption{Percentage histogram of SINR improvement over the subset with small movements of the SOI.}
     \label{fig:hist_small}
\end{subfigure}
  \caption{Histograms of SINR improvement achieved by the variants of Block AuxIVE in the experiment of Section~\ref{sec:mirageexperiment}.}
     \label{fig:hist}
 \end{figure*}

The SOI is blindly extracted from each mixture, and the result is evaluated through the improvement of the Signal-to-Interference-and-Noise ratio (iSINR) and Signal-to-Distortion ratio (iSDR) defined as in \cite{koldovsky2013TASLP} (SDR is computed after compensating for the global delay). The averaged values of the criteria are summarized in Table~\ref{tab:mirage_results} together with the average time to process one mixture.
For a deeper understanding to the results, we also analyze the histograms of iSINR by OverIVA and Block AuxIVA shown in Fig.~\ref{fig:hist}.

Fig.~\ref{fig:hist_full} shows the histograms over full data set of mixtures, while Fig.~\ref{fig:hist_small} is evaluated on a subset of mixtures in which the SOI has not moved away from the starting position by more than $5$~cm; there are $288$ mixtures of this kind. Now, we can observe two phenomena. First, it can be seen that the methods for the static model yield more results below 10~dB in Fig.~\ref{fig:hist_full} than in Fig.~\ref{fig:hist_small}. 
That means that they perform better for the subset of mixtures where the SOI is almost static. The performance of the block-based methods seems to be similar for the full set and the subset. To summarize, the block methods yield a more stable performance than the static methods when the SOI is moving. 
Second, the piloted methods yield iSINR$<-5$~dB in a much lower number of trials than the non-piloted methods, as confirmed by the additional criterion in  Table~\ref{tab:mirage_results}. This shows that the piloted algorithms have significantly improved global convergence. 



\subsection{Speech enhancement/recognition on CHiME-4 datasets}

We have verified the proposed methods also in the noisy speech recognition task defined within the CHiME-4 challenge, specifically, the six-channel track \cite{chime_data_soft}. This dataset contains simulated (SIMU) and real-world\footnote{Microphone $2$ is not used in the case of the real-world recordings as, here, it is oriented away from the speaker.} (REAL) utterances of speakers in multi-source noisy environments. The recording device is a tablet with six microphones, which is held by a speaker.
Since some recordings involve microphone failures, the method from \cite{our_journal_beam} is used to detect these failures. If detected, the malfunctioning channels are excluded from further processing of the given recording.

The experiment is evaluated in terms of Word Error Rate (WER) as follows: The compared methods are used to extract speech from the noisy recordings. Then, the enhanced signals are forwarded to the baseline speech recognizer from \cite{chime_data_soft}. The WER achieved by the proposed methods is compared with the results obtained on unprocessed input signals (Channel $5$) and with the techniques listed below.

BeamformIt~\cite{beamformit} is a front-end algorithm used within the CHiME-4 baseline system. It is a weighted delay-and-sum beamformer requiring two passes over the processed recording in order to optimize its inner parameters. We use the original implementation of the technique available at  \cite{chime_web}.

The Generalized Eigenvalue Beamformer (GEV) is a front-end solution proposed in \cite{heymann, heymann_chime4}. It represents the most successful enhancers for CHiME-4 that rely on deep networks trained for the CHiME-4 data. In the implementation used here, a re-trained Voice-Activity-Detector (VAD) is used where the training procedure was kindly provided by the authors of \cite{heymann}. We utilize the feed-forward topology of the VAD and train the network using the training part of the CHiME-4 data. GEV utilizes the Blind Analytic Normalization (BAN) postfilter to obtain its final enhanced output signal.

All systems/algorithms operate in the STFT domain with an FFT length of $512$, a hop-size of $128$ and use the Hamming window; the sampling frequency is $16$~kHz. BOGIVE$_{\bf w}$ and Block AuxIVE are applied with $N_b=250$, which corresponds to the block length of $2$~s. This value has been selected to optimize the performance of these methods. All of the proposed methods are initialized by the Relative Transfer Function (RTF) estimator from \cite{gannot2001}; Channel 5 of the data is selected as the target (the spatial image of the speech signal of this channel is being estimated).

\begin{table}
\centering
\caption{WERs [\%] achieved in the CHiME-4 challenge. \label{tab:wer}}	
\begin{tabular}{ccccc}
	\multirow{2}{*}{\begin{tabular}[c]{@{}c@{}}System
	\end{tabular}} & \multicolumn{2}{c}{Development} &
	\multicolumn{2}{c}{Test}\\
	\cline{2-5}
	& REAL & SIMU & REAL & SIMU\\
	\hline
	Unprocessed  		& 9.83 & 8.86 & 19.90 & 10.79\\
	BeamformIt   		& 5.77 & 6.76 & 11.52 & 10.91\\
	GEV (VAD) 			& \textbf{4.61} & \textbf{4.65} & \textbf{8.10}  & \textbf{5.99}\\		
    OGIVE$_{\bf w}$ & 5.59 & 4.96 & 9.51 & 6.34\\
	BOGIVE$_{\bf w}$   & 5.49 & 4.91 & 9.19 & 6.44\\
	OverIVA & 5.97 & 5.21 & 10.43 & 6.82 \\
    Block AuxIVE & 5.65 & 4.83 & 9.88 & 6.46\\		
\end{tabular}
\end{table}

The results shown in Table~\ref{tab:wer} indicate that all methods are able to improve the WER compared to the unprocessed case. The BSE-based methods significantly outperform BeamformIt. The GEV beamformer endowed with the pretrained VAD achieves the best results. It should be noted that the rates achieved by the BSE techniques are comparable to GEV even without a training stage on any CHiME-4 data. 

In general, the block-wise methods achieve lower WER than their counterparts based on the static mixing model; the WER of BOGIVE$_{\bf w}$ is comparable with Block AuxIVE. A significant advantage of the latter method is the faster convergence and, consequently, much lower computational burden. The total duration of the $5920$ files in the CHiME-4 dataset is $10$~hours and $5$~minutes. The results presented for BOGIVE$_{\bf w}$ have been achieved after $100$ iterations on each file, which translates into $10$~hours and $30$ minutes\footnote{The computations run on a workstation using an Intel i7-2600K@3.4GHz processor with 16GB RAM.} of processing for the whole dataset. Block AuxIVE is able to converge in $7$ iterations; the whole enhancement was finished in $1$ hour and $2$ minutes.

\begin{figure}
 \centering
 \includegraphics[width=0.99\linewidth]{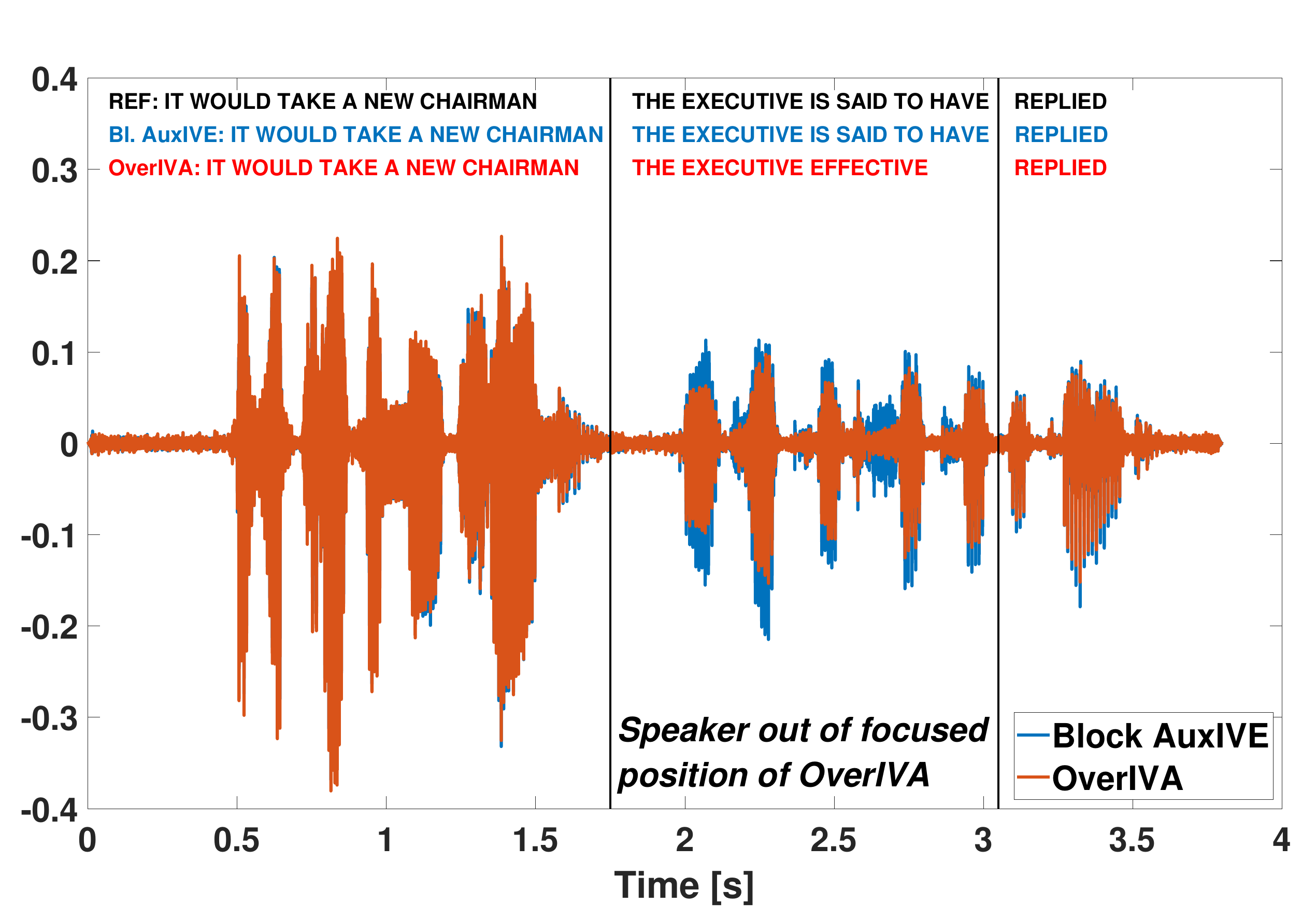}
 \caption{\label{fig:casestudy} Comparison of enhanced signals yielded from a recording of a moving speaker by OverIVA and Block AuxIVE.}
\end{figure}

An example of the enhancement yielded by the block-wise methods on one of the CHiME-4 recordings is shown in Fig.~\ref{fig:casestudy}.
Within this particular recording, in the interval $1.75-3$~s, the target speaker was moved out of its initial position. The OverIVA algorithm focused on this initial direction only, resulting in vanishing voice during the movement interval.
Consequently, the automatic transcription is erroneous. In contrast, Block AuxIVE is able to focus on both positions of the speaker and recovers the signal of interest correctly. The fact that there are few such recordings with significant speaker movement in the CHiME-4 datasets explains why the achieved improvements of WER by the block-wise methods are small.

\section{Conclusions}
The ability of the CSV-based BSE algorithms to extract moving acoustic sources has been corroborated by the experiments presented in this paper. The blind extraction is based on the estimation of a separating filter that passes signals from the entire area of the source presence. This way, the moving source can be extracted efficiently without tracking in an on-line fashion. The experiments show that these methods are particularly robust with respect to small source movements and effectively exploit overdetermined settings, that is, when there is a higher number of microphones than that of the sources.

We have proposed a new BSE algorithm of this kind, Block AuxIVE, which is based on the auxiliary function-based optimization. The algorithm was shown to be faster in convergence compared to its gradient-based counterpart. 
Furthermore, we have proposed the semi-supervised variant of Block AuxIVE utilizing pilot signals. The experiments confirm that this algorithm yields stable global convergence to the SOI even when the pilot signal is only a roughly pre-extracted SOI containing a considerable residual of noise and interference. 

For the future, the proposed methods provide us with alternatives to the conventional approaches that adapt to the source movements through application of static mixing models on short time-intervals. Their other abilities, for example the robustness against a highly reverberant and noisy environment, poses an interesting topic for future research; see \cite{malek2021}.



\input{main.bbl}

\end{document}